\documentclass[11pt]{article}
\pdfoutput=1

\usepackage{jheppub}
\usepackage[utf8]{inputenc}
\usepackage{booktabs}
\usepackage{graphicx}
\usepackage{bbold}
\usepackage{subcaption}
\usepackage{mathtools}

\input{widebar}
\usepackage{xkeyval}

\usepackage{tikz,fp}
\usetikzlibrary{external,fixedpointarithmetic,decorations.pathmorphing,positioning,calc,fadings,decorations.markings,decorations.pathreplacing,patterns,arrows}

\tikzset{
  label distance=-3pt,
  >=stealth,
  inner sep=1.5pt,
  boson/.style={
    decoration={snake, segment length=2mm, amplitude=0.5mm},
    decorate,
  },
  quark/.style={
    postaction={decorate},
    decoration={markings,mark=at position .5 with {\draw[->] (-0.01,0) -- (0.07,0);}}
  },
  aquark/.style={
    postaction={decorate},
    decoration={markings,mark=at position .5 with {\draw[->] (0.01,0) -- (-0.07,0);}}
  },
  gluonOld/.style={
    line width=0.7pt,
    decorate,
    decoration={coil,aspect=-0.5,amplitude=-2pt,segment length=2.2pt}
  },
  gluon/.style={
    line width=0.5pt,
    decorate,
    decoration={coil,aspect=-0.75,amplitude=-1.5pt,segment length=2pt}
  },
  agluon/.style={
    line width=0.5pt,
    decorate,
    decoration={coil,aspect=0.75,amplitude=1.5pt,segment length=2pt}
  },
  scalar/.style={
    dashed
  },
  dscalar/.style={
    dashed,
    postaction={decorate},
    decoration={markings,mark=at position 0.65 with {\draw[->] (-0.01,0) -- (0.07,0);}}
  },
  adscalar/.style={
    dashed,
    postaction={decorate},
    decoration={markings,mark=at position 0.65 with {\draw[->] (-0.01,0) -- (0.07,0);}}
  },
  line/.style={
  },
  doubleline/.style={
    double
  },
  middleArrow/.style={
    draw=white,
    decoration={markings,mark=at position 0.65 with{\arrow[scale=1,black]{>}}},
    decorate
  },
  dot/.style={
  	postaction={decorate},
    decoration={markings,mark=between positions 0.5 and 0.5 step 1 with {\draw[fill] (0,0) circle [radius=1.5pt];}}  
  },
  dots/.style={
  	postaction={decorate},
    decoration={markings,mark=between positions 0.3 and 0.6 step 0.3 with {\draw[fill] (0,0) circle [radius=1.5pt];}}  
  },
  dotss/.style={
  	postaction={decorate},
    decoration={markings,mark=between positions 0.25 and 0.75 step 0.25 with {\draw[fill] (0,0) circle [radius=1.5pt];}}  
  },
  blob/.style={
  	pattern=north west lines,
    draw=black
  },
  blobBlack/.style={
    draw=black,
    fill=black
  },
  blobWhite/.style={
    draw=black,
    fill=white
  }
}

\makeatletter
\def\gSetStdExtTypes#1{\presetkeys{gTypes}{eA=#1,eB=#1,eC=#1,eD=#1,eE=#1}{}}
\def\gSetStdIntTypes#1{\presetkeys{gTypes}{iA=#1,iB=#1,iC=#1,iD=#1,iE=#1,iF=#1,iG=#1,iH=#1,iI=#1,iJ=#1}{}}
\def\gSetStdTypes#1{\gSetStdExtTypes{#1}\gSetStdIntTypes{#1}}

\define@cmdkeys{general}{scale}
\define@cmdkeys{general}{rotate}
\define@cmdkeys{general}{font}
\define@cmdkeys{general}{yshift}
\presetkeys{general}{scale=1, rotate=0, font=\scriptsize, yshift=0pt}{}

\define@cmdkeys{gTypes}{eA,eB,eC,eD,eE,iA,iB,iC,iD,iE,iF,iG,iH,iI,iJ}
\gSetStdTypes{line}
\define@key{pre}{all}{\gSetStdTypes{#1}}
\define@key{pre}{allE}{\gSetStdExtTypes{#1}}
\define@key{pre}{allI}{\gSetStdIntTypes{#1}}

\define@cmdkeys{gLabels}{eLA,eLB,eLC,eLD,eLE,iLA,iLB,iLC,iLD,iLE,iLF,iLG,iLH,iLI,iLJ}
\presetkeys{gLabels}{eLA={},eLB={},eLC={},eLD={},eLE={},iLA={},iLB={},iLC={},iLD={},iLE={},iLF={},iLG={},iLH={},iLI={},iLJ={}}{}

\define@boolkeys{dots}{lDots,rDots}
\presetkeys{dots}{lDots=false, rDots=false}{}

\define@cmdkeys{blobs}{blobA,blobB,blobC}
\presetkeys{blobs}{blobA=blob,blobB=blob,blobC=blob}{}

\def\gTreeTri{\@ifnextchar[\@gTreeTri{\@gTreeTri[]}}
\def\@gTreeTri[#1]#2{%
  \setkeys*{pre}{#1}%
  \setrmkeys{general,gTypes,gLabels}%
  \begin{tikzpicture}
    [line width=1pt,
    baseline={([yshift=-0.5ex+\cmdKV@general@yshift]current bounding box.center)},
    scale=\cmdKV@general@scale,
    rotate=\cmdKV@general@rotate,
    font=\cmdKV@general@font]
    \draw[\cmdKV@gTypes@eA] (0.7,0.3) -- (0.4,0.3) node[label=(-180+\cmdKV@general@rotate):\cmdKV@gLabels@eLA] {};
    \draw[\cmdKV@gTypes@eB] (0.7,0.3) -- (0.9,0.6) node[label=(\cmdKV@general@rotate):\cmdKV@gLabels@eLB] {};
    \draw[\cmdKV@gTypes@eC] (0.7,0.3) -- (0.9,0) node[label=(\cmdKV@general@rotate):\cmdKV@gLabels@eLC] {};
    #2
  \end{tikzpicture}
  \gSetStdTypes{line}
}

\def\gTreeS{\@ifnextchar[\@gTreeS{\@gTreeS[]}}
\def\@gTreeS[#1]#2{%
  \setkeys*{pre}{#1}%
  \setrmkeys{general,gTypes,gLabels}%
  \begin{tikzpicture}
    [line width=1pt,
    baseline={([yshift=-0.5ex+\cmdKV@general@yshift]current bounding box.center)},
    scale=\cmdKV@general@scale,
    rotate=\cmdKV@general@rotate,
    font=\cmdKV@general@font]
    \draw[\cmdKV@gTypes@eA] (0.2,0.3) -- (0,0) node[label=(-180+\cmdKV@general@rotate):\cmdKV@gLabels@eLA] {};
    \draw[\cmdKV@gTypes@eB] (0.2,0.3) -- (0,0.6) node[label=(180+\cmdKV@general@rotate):\cmdKV@gLabels@eLB] {};
    \draw[\cmdKV@gTypes@eC] (0.7,0.3) -- (0.9,0.6) node[label=(\cmdKV@general@rotate):\cmdKV@gLabels@eLC] {};
    \draw[\cmdKV@gTypes@eD] (0.7,0.3) -- (0.9,0) node[label=(\cmdKV@general@rotate):\cmdKV@gLabels@eLD] {};    
    \draw[label distance=0,\cmdKV@gTypes@iA] (0.2,0.3) -- node[label=(90+\cmdKV@general@rotate):\cmdKV@gLabels@iLA] {} (0.7,0.3);
    #2
  \end{tikzpicture}
  \gSetStdTypes{line}
}

\def\gTreeT{\@ifnextchar[\@gTreeT{\@gTreeT[]}}
\def\@gTreeT[#1]#2{%
  \setkeys*{pre}{#1}%
  \setrmkeys{general,gTypes,gLabels}%
  \begin{tikzpicture}
    [line width=1pt,
    baseline={([yshift=-0.5ex+\cmdKV@general@yshift]current bounding box.center)},
    scale=\cmdKV@general@scale,
    rotate=\cmdKV@general@rotate,
    font=\cmdKV@general@font]
    \draw[\cmdKV@gTypes@eA] (0.45,0.1) -- (0,0) node[label=(-180+\cmdKV@general@rotate):\cmdKV@gLabels@eLA] {};
    \draw[\cmdKV@gTypes@eB] (0.45,0.5) -- (0,0.6) node[label=(180+\cmdKV@general@rotate):\cmdKV@gLabels@eLB] {};
    \draw[\cmdKV@gTypes@eC] (0.45,0.5) -- (0.9,0.6) node[label=(\cmdKV@general@rotate):\cmdKV@gLabels@eLC] {};
    \draw[\cmdKV@gTypes@eD] (0.45,0.1) -- (0.9,0) node[label=(\cmdKV@general@rotate):\cmdKV@gLabels@eLD] {};
    \draw[label distance=0,\cmdKV@gTypes@iA] (0.45,0.5) -- node[label=(\cmdKV@general@rotate):\cmdKV@gLabels@iLA] {} (0.45,0.1);
    #2
  \end{tikzpicture}
  \gSetStdTypes{line}
}

\def\gTreeU{\@ifnextchar[\@gTreeU{\@gTreeU[]}}
\def\@gTreeU[#1]#2{%
  \setkeys*{pre}{#1}%
  \setrmkeys{general,gTypes,gLabels}%
  \begin{tikzpicture}
    [line width=1pt,
    baseline={([yshift=-0.5ex+\cmdKV@general@yshift]current bounding box.center)},
    scale=\cmdKV@general@scale,
    rotate=\cmdKV@general@rotate,
    font=\cmdKV@general@font]
    \draw[\cmdKV@gTypes@eA] (0.45,0.5) -- (0,0) node[label=(-180+\cmdKV@general@rotate):\cmdKV@gLabels@eLA] {};
    \foreach \x in {5,...,14} \fill[opacity=0.12,white] (0.27,0.3) circle (\x/100);
    \draw[\cmdKV@gTypes@eB] (0.45,0.1) -- (0,0.6) node[label=(180+\cmdKV@general@rotate):\cmdKV@gLabels@eLB] {};
    \draw[\cmdKV@gTypes@eC] (0.45,0.5) -- (0.9,0.6) node[label=(\cmdKV@general@rotate):\cmdKV@gLabels@eLC] {};
    \draw[\cmdKV@gTypes@eD] (0.45,0.1) -- (0.9,0) node[label=(\cmdKV@general@rotate):\cmdKV@gLabels@eLD] {};
    \draw[label distance=0,\cmdKV@gTypes@iA] (0.45,0.5) -- node[label=(\cmdKV@general@rotate):\cmdKV@gLabels@iLA] {} (0.45,0.1);
    #2
  \end{tikzpicture}
  \gSetStdTypes{line}
}

\def\gTreeFourPt{\@ifnextchar[\@gTreeFourPt{\@gTreeFourPt[]}}
\def\@gTreeFourPt[#1]#2{%
  \setkeys*{pre}{#1}%
  \setrmkeys{general,gTypes,gLabels}%
  \begin{tikzpicture}
    [line width=1pt,
    baseline={([yshift=-0.5ex+\cmdKV@general@yshift]current bounding box.center)},
    scale=\cmdKV@general@scale,
    rotate=\cmdKV@general@rotate,
    font=\cmdKV@general@font]
    \draw[\cmdKV@gTypes@eA] (0,0) -- (-0.4,-0.4) node[label=(-180+\cmdKV@general@rotate):\cmdKV@gLabels@eLA] {};
    \draw[\cmdKV@gTypes@eB] (0,0) -- (-0.4,0.4) node[label=(180+\cmdKV@general@rotate):\cmdKV@gLabels@eLB] {};
    \draw[\cmdKV@gTypes@eC] (0,0) -- (0.4,0.4) node[label=(\cmdKV@general@rotate):\cmdKV@gLabels@eLC] {};
    \draw[\cmdKV@gTypes@eD] (0,0) -- (0.4,-0.4) node[label=(\cmdKV@general@rotate):\cmdKV@gLabels@eLD] {};
    #2
  \end{tikzpicture}
  \gSetStdTypes{line}
}

\def\gTadA{\@ifnextchar[\@gTadA{\@gTadA[]}}
\def\@gTadA[#1]#2{%
  \setkeys*{pre}{#1}%
  \setrmkeys{general,gTypes,gLabels}%
  \begin{tikzpicture}
    [line width=1pt,
    baseline={([yshift=-0.5ex+\cmdKV@general@yshift]current bounding box.center)},
    scale=\cmdKV@general@scale,
    rotate=\cmdKV@general@rotate,
    font=\cmdKV@general@font]
    \draw[\cmdKV@gTypes@eA] (-0.7,0.3) -- (-0.9,0) node[label=(180+\cmdKV@general@rotate):\cmdKV@gLabels@eLA] {};
    \draw[\cmdKV@gTypes@eB] (-0.7,0.3) -- (-0.9,0.6) node[label=(180+\cmdKV@general@rotate):\cmdKV@gLabels@eLB] {};
    \draw[\cmdKV@gTypes@eC] (0.15,0.45) -- (0.5,0.6) node[label=(\cmdKV@general@rotate):\cmdKV@gLabels@eLC] {};
    \draw[\cmdKV@gTypes@eD] (-0.3,0.3) -- (-0.2,0) node[label=(\cmdKV@general@rotate):\cmdKV@gLabels@eLD] {};
    \draw[label distance=0,\cmdKV@gTypes@iA] (-0.7,0.3) -- node[label=(90+\cmdKV@general@rotate):\cmdKV@gLabels@iLA] {} (-0.3,0.3);
    \draw[label distance=0,\cmdKV@gTypes@iB] (-0.3,0.3) -- node[label=(-80+\cmdKV@general@rotate):\cmdKV@gLabels@iLB] {} (0.15,0.45);
    \draw[label distance=0,\cmdKV@gTypes@iC] (0.15,0.45) -- node[label=(45+\cmdKV@general@rotate):\cmdKV@gLabels@iLC] {} (-0.18,0.69);
    \draw[label distance=-0.5,\cmdKV@gTypes@iD] (-0.18,0.69) .. controls ($(-0.18,0.69) + (-0.12,-0.12)$) and ($(-0.35,0.87) + (-0.12,-0.12)$) .. (-0.35,0.87) node[label=(180+\cmdKV@general@rotate):\cmdKV@gLabels@iLD] {} .. controls ($(-0.35,0.87) + (0.12,0.12)$) and ($(-0.18,0.69) + (0.12,0.12)$) .. (-0.18,0.69);
    #2
  \end{tikzpicture}
  \gSetStdTypes{line}
}

\def\gTadB{\@ifnextchar[\@gTadB{\@gTadB[]}}
\def\@gTadB[#1]#2{%
  \setkeys*{pre}{#1}%
  \setrmkeys{general,gTypes,gLabels}%
  \begin{tikzpicture}
    [line width=1pt,
    baseline={([yshift=-0.5ex+\cmdKV@general@yshift]current bounding box.center)},
    scale=\cmdKV@general@scale,
    rotate=\cmdKV@general@rotate,
    font=\cmdKV@general@font]
    \draw[\cmdKV@gTypes@eA] (0,0.3) -- (-0.2,0) node[label=(-180+\cmdKV@general@rotate):\cmdKV@gLabels@eLA] {};
    \draw[\cmdKV@gTypes@eB] (0,0.3) -- (-0.2,0.6) node[label=(180+\cmdKV@general@rotate):\cmdKV@gLabels@eLB] {};
    \draw[\cmdKV@gTypes@eC] (0.7,0.3) -- (0.9,0.6) node[label=(\cmdKV@general@rotate):\cmdKV@gLabels@eLC] {};
    \draw[\cmdKV@gTypes@eD] (0.7,0.3) -- (0.9,0) node[label=(\cmdKV@general@rotate):\cmdKV@gLabels@eLD] {};
    \draw[label distance=0,\cmdKV@gTypes@iA] (0,0.3) -- node[label=(-90+\cmdKV@general@rotate):\cmdKV@gLabels@iLA] {} (0.35,0.3);
    \draw[label distance=0,\cmdKV@gTypes@iB] (0.35,0.3) -- node[label=(\cmdKV@general@rotate):\cmdKV@gLabels@iLB] {} (0.35,0.6);
    \draw[label distance=-0.5,\cmdKV@gTypes@iC] (0.35,0.6) .. controls (0.18,0.6) and (0.18,0.85) .. (0.35,0.85) .. controls (0.52,0.85) and (0.52,0.6) .. node[label=(\cmdKV@general@rotate):\cmdKV@gLabels@iLC] {} (0.35,0.6);
    \draw[label distance=0,\cmdKV@gTypes@iD] (0.35,0.3) -- node[label=(-90+\cmdKV@general@rotate):\cmdKV@gLabels@iLD] {} (0.7,0.3);
    #2
  \end{tikzpicture}
  \gSetStdTypes{line}
}
\def\gBubA{\@ifnextchar[\@gBubA{\@gBubA[]}}
\def\@gBubA[#1]#2{%
  \setkeys*{pre}{#1}%
  \setrmkeys{general,gTypes,gLabels}%
  \begin{tikzpicture}
    [line width=1pt,
    baseline={([yshift=-0.5ex+\cmdKV@general@yshift]current bounding box.center)},
    scale=\cmdKV@general@scale,
    rotate=\cmdKV@general@rotate,
    font=\cmdKV@general@font]
    \draw[\cmdKV@gTypes@eA] (0.5,0.7) -- (0.2,0.7) node[label=(180+\cmdKV@general@rotate):\cmdKV@gLabels@eLA] {};
    \draw[\cmdKV@gTypes@eB] (0.9,0.7) -- (1.2,0.7) node[label=(\cmdKV@general@rotate):\cmdKV@gLabels@eLB] {};
    \draw[label distance=0,\cmdKV@gTypes@iA] (0.5,0.7) .. controls (0.5,0.95) and (0.9,0.95) .. node[label=(90+\cmdKV@general@rotate):\cmdKV@gLabels@iLA] {} (0.9,0.7);
    \draw[label distance=0,\cmdKV@gTypes@iB] (0.9,0.7) .. controls (0.9,0.45) and (0.5,0.45) .. node[label=(-90+\cmdKV@general@rotate):\cmdKV@gLabels@iLB] {} (0.5,0.7);
    #2
  \end{tikzpicture}
  \gSetStdTypes{line}
}
\def\gBubB{\@ifnextchar[\@gBubB{\@gBubB[]}}
\def\@gBubB[#1]#2{%
  \setkeys*{pre}{#1}%
  \setrmkeys{general,gTypes,gLabels}%
  \begin{tikzpicture}
    [line width=1pt,
    baseline={([yshift=-0.5ex+\cmdKV@general@yshift]current bounding box.center)},
    scale=\cmdKV@general@scale,
    rotate=\cmdKV@general@rotate,
    font=\cmdKV@general@font]
    \draw[\cmdKV@gTypes@eA] (0.2,0.7) -- (0,0.4) node[label=(-180+\cmdKV@general@rotate):\cmdKV@gLabels@eLA] {};
    \draw[\cmdKV@gTypes@eB] (0.2,0.7) -- (0,1) node[label=(180+\cmdKV@general@rotate):\cmdKV@gLabels@eLB] {};
    \draw[\cmdKV@gTypes@eC] (1.2,0.7) -- (1.4,1) node[label=(\cmdKV@general@rotate):\cmdKV@gLabels@eLC] {};
    \draw[\cmdKV@gTypes@eD] (1.2,0.7) -- (1.4,0.4) node[label=(\cmdKV@general@rotate):\cmdKV@gLabels@eLD] {};
    \draw[label distance=0,\cmdKV@gTypes@iA] (0.2,0.7) -- node[label=(90+\cmdKV@general@rotate):\cmdKV@gLabels@iLA] {} (0.5,0.7);
    \draw[label distance=0,\cmdKV@gTypes@iB] (0.5,0.7) .. controls (0.5,0.95) and (0.9,0.95) .. node[label=(90+\cmdKV@general@rotate):\cmdKV@gLabels@iLB] {} (0.9,0.7);
    \draw[label distance=0,\cmdKV@gTypes@iC] (0.9,0.7) .. controls (0.9,0.45) and (0.5,0.45) .. node[label=(-90+\cmdKV@general@rotate):\cmdKV@gLabels@iLC] {} (0.5,0.7);
    \draw[label distance=0,\cmdKV@gTypes@iD] (0.9,0.7) -- node[label=(90+\cmdKV@general@rotate):\cmdKV@gLabels@iLD] {} (1.2,0.7);
    #2
  \end{tikzpicture}
  \gSetStdTypes{line}
}

\def\gBubC{\@ifnextchar[\@gBubC{\@gBubC[]}}
\def\@gBubC[#1]#2{%
  \setkeys*{pre}{#1}%
  \setrmkeys{general,gTypes,gLabels}%
  \begin{tikzpicture}
    [line width=1pt,
    baseline={([yshift=-0.5ex+\cmdKV@general@yshift]current bounding box.center)},
    scale=\cmdKV@general@scale,
    rotate=\cmdKV@general@rotate,
    font=\cmdKV@general@font]
    \draw[\cmdKV@gTypes@eA] (-0.7,0.3) -- (-0.9,0) node[label=(180+\cmdKV@general@rotate):\cmdKV@gLabels@eLA] {};
    \draw[\cmdKV@gTypes@eB] (-0.7,0.3) -- (-0.9,0.6) node[label=(180+\cmdKV@general@rotate):\cmdKV@gLabels@eLB] {};
    \draw[\cmdKV@gTypes@eC] (0.3,0.5) -- (0.5,0.6) node[label=(\cmdKV@general@rotate):\cmdKV@gLabels@eLC] {};
    \draw[\cmdKV@gTypes@eD] (-0.3,0.3) -- (-0.2,0) node[label=(\cmdKV@general@rotate):\cmdKV@gLabels@eLD] {};
    \draw[label distance=0,\cmdKV@gTypes@iA] (-0.7,0.3) -- node[label=(90+\cmdKV@general@rotate):\cmdKV@gLabels@iLA] {} (-0.3,0.3);
    \draw[label distance=0,\cmdKV@gTypes@iB] (-0.3,0.3) -- node[label=(90+\cmdKV@general@rotate):\cmdKV@gLabels@iLB] {} (-0,0.4);
    \draw[label distance=0,\cmdKV@gTypes@iC] (0,0.4) .. controls (-0.03,0.6) and (0.22,0.65) .. node[label=(90+\cmdKV@general@rotate):\cmdKV@gLabels@iLC] {} (0.3,0.5);
    \draw[label distance=0,\cmdKV@gTypes@iD] (0.3,0.5) .. controls (0.38,0.3) and (0.08,0.21) .. node[label=(-90+\cmdKV@general@rotate):\cmdKV@gLabels@iLD] {} (0,0.4);
    #2
  \end{tikzpicture}
  \gSetStdTypes{line}
}
\def\gTriA{\@ifnextchar[\@gTriA{\@gTriA[]}}
\def\@gTriA[#1]#2{%
  \setkeys*{pre}{#1}%
  \setrmkeys{general,gTypes,gLabels}%
  \begin{tikzpicture}
    [line width=1pt,
    baseline={([yshift=-0.5ex+\cmdKV@general@yshift]current bounding box.center)},
    scale=\cmdKV@general@scale,
    rotate=\cmdKV@general@rotate,
    font=\cmdKV@general@font]
    \draw[\cmdKV@gTypes@eC] (0:0.3) -- (0:0.55) node[label=(\cmdKV@general@rotate):\cmdKV@gLabels@eLC] {};
    \draw[\cmdKV@gTypes@eA] (-120:0.3) -- (-120:0.55) node[label=(-180+\cmdKV@general@rotate):\cmdKV@gLabels@eLA] {};
    \draw[\cmdKV@gTypes@eB] (120:0.3) -- (120:0.55) node[label=(180+\cmdKV@general@rotate):\cmdKV@gLabels@eLB] {};    
    \draw[label distance=0,\cmdKV@gTypes@iA] (-120:0.3) -- node[label=(180+\cmdKV@general@rotate):\cmdKV@gLabels@iLA] {} (120:0.3);
    \draw[label distance=0,\cmdKV@gTypes@iB] (120:0.3) -- node[label=(60+\cmdKV@general@rotate):\cmdKV@gLabels@iLB] {} (0:0.3);
    \draw[label distance=0,\cmdKV@gTypes@iC] (0:0.3) -- node[label=(-60+\cmdKV@general@rotate):\cmdKV@gLabels@iLC] {} (-120:0.3);
    #2
  \end{tikzpicture}
  \gSetStdTypes{line}
}
\def\gTriB{\@ifnextchar[\@gTriB{\@gTriB[]}}
\def\@gTriB[#1]#2{%
  \setkeys*{pre}{#1}%
  \setrmkeys{general,gTypes,gLabels}%
  \begin{tikzpicture}
    [line width=1pt,
    baseline={([yshift=-0.5ex+\cmdKV@general@yshift]current bounding box.center)},
    scale=\cmdKV@general@scale,
    rotate=\cmdKV@general@rotate,
    font=\cmdKV@general@font]
    \draw[\cmdKV@gTypes@eA] (0.3,0.45) -- (0,0.3) node[label=(-180+\cmdKV@general@rotate):\cmdKV@gLabels@eLA] {};
    \draw[\cmdKV@gTypes@eB] (0.3,0.95) -- (0,1.1) node[label=(180+\cmdKV@general@rotate):\cmdKV@gLabels@eLB] {};
    \draw[\cmdKV@gTypes@eC] (1,0.7) -- (1.3,1.1) node[label=(\cmdKV@general@rotate):\cmdKV@gLabels@eLC] {};
    \draw[\cmdKV@gTypes@eD] (1,0.7) -- (1.3,0.3) node[label=(\cmdKV@general@rotate):\cmdKV@gLabels@eLD] {};
    \draw[label distance=0,\cmdKV@gTypes@iA] (0.3,0.45) -- node[label=(180+\cmdKV@general@rotate):\cmdKV@gLabels@iLA] {} (0.3,0.95);
    \draw[label distance=0,\cmdKV@gTypes@iB] (0.3,0.95) -- node[label=(70+\cmdKV@general@rotate):\cmdKV@gLabels@iLB] {} (0.7,0.7);
    \draw[label distance=0,\cmdKV@gTypes@iC] (0.7,0.7) -- node[label=(-70+\cmdKV@general@rotate):\cmdKV@gLabels@iLC] {} (0.3,0.45);
    \draw[label distance=0,\cmdKV@gTypes@iD] (1,0.7) -- node[label=(90+\cmdKV@general@rotate):\cmdKV@gLabels@iLD] {} (0.7,0.7);
    #2
  \end{tikzpicture}
  \gSetStdTypes{line}
}
\def\gBox{\@ifnextchar[\@gBox{\@gBox[]}}
\def\@gBox[#1]#2{%
  \setkeys*{pre}{#1}%
  \setrmkeys{general,gTypes,gLabels}%
  \begin{tikzpicture}
    [line width=1pt,
    baseline={([yshift=-0.5ex+\cmdKV@general@yshift]current bounding box.center)},
    scale=\cmdKV@general@scale,
    rotate=\cmdKV@general@rotate,
    font=\cmdKV@general@font]
    \draw[\cmdKV@gTypes@eA] (0.25,0.25) -- (0,0) node[label=(-180+\cmdKV@general@rotate):\cmdKV@gLabels@eLA] {};
    \draw[\cmdKV@gTypes@eB] (0.25,0.75) -- (0,1) node[label=(180+\cmdKV@general@rotate):\cmdKV@gLabels@eLB] {};
    \draw[\cmdKV@gTypes@eC] (0.75,0.75) -- (1,1) node[label=(\cmdKV@general@rotate):\cmdKV@gLabels@eLC] {};
    \draw[\cmdKV@gTypes@eD] (0.75,0.25) -- (1,0) node[label=(\cmdKV@general@rotate):\cmdKV@gLabels@eLD] {};
    \draw[label distance=0,\cmdKV@gTypes@iA] (0.25,0.25) -- node[label=(180+\cmdKV@general@rotate):\cmdKV@gLabels@iLA] {} (0.25,0.75);
    \draw[label distance=0,\cmdKV@gTypes@iB] (0.25,0.75) -- node[label=(90+\cmdKV@general@rotate):\cmdKV@gLabels@iLB] {} (0.75,0.75);
    \draw[label distance=0,\cmdKV@gTypes@iC] (0.75,0.75) -- node[label=(\cmdKV@general@rotate):\cmdKV@gLabels@iLC] {} (0.75,0.25);
    \draw[label distance=0,\cmdKV@gTypes@iD] (0.75,0.25) -- node[label=(-90+\cmdKV@general@rotate):\cmdKV@gLabels@iLD] {} (0.25,0.25);
    #2
  \end{tikzpicture}
  \gSetStdTypes{line}
}
\def\gPenta{\@ifnextchar[\@gPenta{\@gPenta[]}}
\def\@gPenta[#1]#2{%
  \setkeys*{pre}{#1}%
  \setrmkeys{general,gTypes,gLabels}%
  \begin{tikzpicture}
    [line width=1pt,
    baseline={([yshift=-0.5ex+\cmdKV@general@yshift]current bounding box.center)},
    scale=\cmdKV@general@scale,
    rotate=\cmdKV@general@rotate,
    font=\cmdKV@general@font]
    \draw[\cmdKV@gTypes@eA] (-144:0.35) -- (-144:0.75) node[label=(-144+\cmdKV@general@rotate):\cmdKV@gLabels@eLA] {};
    \draw[\cmdKV@gTypes@eB] (144:0.35) -- (144:0.75) node[label=(144+\cmdKV@general@rotate):\cmdKV@gLabels@eLB] {};
    \draw[\cmdKV@gTypes@eC] (72:0.35) -- (72:0.75) node[label=(72+\cmdKV@general@rotate):\cmdKV@gLabels@eLC] {};
    \draw[\cmdKV@gTypes@eD] (0:0.35) -- (0:0.75) node[label=(\cmdKV@general@rotate):\cmdKV@gLabels@eLD] {};
    \draw[\cmdKV@gTypes@eE] (-72:0.35) -- (-72:0.75) node[label=(-72+\cmdKV@general@rotate):\cmdKV@gLabels@eLE] {};
    \draw[label distance=0,\cmdKV@gTypes@iA] (-144:0.35) -- node[label=(180+\cmdKV@general@rotate):\cmdKV@gLabels@iLA] {} (144:0.35);
    \draw[label distance=0,\cmdKV@gTypes@iB] (144:0.35) -- node[label=(108+\cmdKV@general@rotate):\cmdKV@gLabels@iLB] {} (72:0.35);
    \draw[label distance=0,\cmdKV@gTypes@iC] (72:0.35) -- node[label=(36+\cmdKV@general@rotate):\cmdKV@gLabels@iLC] {} (0:0.35);
    \draw[label distance=0,\cmdKV@gTypes@iD] (0:0.35) -- node[label=(-36+\cmdKV@general@rotate):\cmdKV@gLabels@iLD] {} (-72:0.35);
    \draw[label distance=0,\cmdKV@gTypes@iE] (-72:0.35) -- node[label=(-108+\cmdKV@general@rotate):\cmdKV@gLabels@iLE] {} (-144:0.35);
    #2
  \end{tikzpicture}
  \gSetStdTypes{line}
}

\def\gBoxBox{\@ifnextchar[\@gBoxBox{\@gBoxBox[]}}
\def\@gBoxBox[#1]#2{%
  \setkeys*{pre}{#1}%
  \setrmkeys{general,gTypes,gLabels}%
  \begin{tikzpicture}
    [line width=1pt,
    baseline={([yshift=-0.5ex+\cmdKV@general@yshift]current bounding box.center)},
    scale=\cmdKV@general@scale,
    rotate=\cmdKV@general@rotate,
    font=\cmdKV@general@font]
    \draw[\cmdKV@gTypes@eA] (0.3,0.3) -- (0,0) node[label=(-180+\cmdKV@general@rotate):\cmdKV@gLabels@eLA] {};
    \draw[\cmdKV@gTypes@eB] (0.3,0.9) -- (0,1.2) node[label=(180+\cmdKV@general@rotate):\cmdKV@gLabels@eLB] {};
    \draw[\cmdKV@gTypes@eC] (1.5,0.9) -- (1.8,1.2) node[label=(\cmdKV@general@rotate):\cmdKV@gLabels@eLC] {};
    \draw[\cmdKV@gTypes@eD] (1.5,0.3) -- (1.8,0) node[label=(\cmdKV@general@rotate):\cmdKV@gLabels@eLD] {};
    \draw[label distance=0,\cmdKV@gTypes@iA] (0.3,0.3) -- node[label=(180+\cmdKV@general@rotate):\cmdKV@gLabels@iLA] {} (0.3,0.9);
    \draw[label distance=0,\cmdKV@gTypes@iB] (0.3,0.9) -- node[label=(90+\cmdKV@general@rotate):\cmdKV@gLabels@iLB] {} (0.9,0.9);
    \draw[label distance=0,\cmdKV@gTypes@iC] (0.9,0.9) -- node[label=(90+\cmdKV@general@rotate):\cmdKV@gLabels@iLC] {} (1.5,0.9);
    \draw[label distance=0,\cmdKV@gTypes@iD] (1.5,0.9) -- node[label=(\cmdKV@general@rotate):\cmdKV@gLabels@iLD] {} (1.5,0.3);
    \draw[label distance=0,\cmdKV@gTypes@iE] (1.5,0.3) -- node[label=(-90+\cmdKV@general@rotate):\cmdKV@gLabels@iLE] {} (0.9,0.3);
    \draw[label distance=0,\cmdKV@gTypes@iF] (0.9,0.3) -- node[label=(-90+\cmdKV@general@rotate):\cmdKV@gLabels@iLF] {} (0.3,0.3);
    \draw[label distance=0,\cmdKV@gTypes@iG] (0.9,0.9) -- node[label=(\cmdKV@general@rotate):\cmdKV@gLabels@iLG] {} (0.9,0.3);
    #2
  \end{tikzpicture}
  \gSetStdTypes{line}
}
\def\gBoxBoxNP{\@ifnextchar[\@gBoxBoxNP{\@gBoxBoxNP[]}}
\def\@gBoxBoxNP[#1]#2{%
  \setkeys*{pre}{#1}%
  \setrmkeys{general,gTypes,gLabels}%
  \begin{tikzpicture}
    [line width=1pt,
    baseline={([yshift=-0.5ex+\cmdKV@general@yshift]current bounding box.center)},
    scale=\cmdKV@general@scale,
    rotate=\cmdKV@general@rotate,
    font=\cmdKV@general@font]
    \draw[\cmdKV@gTypes@eA] (-1.9,0.3) -- (-2.2,0) node[label=(180+\cmdKV@general@rotate):\cmdKV@gLabels@eLA] {};
    \draw[\cmdKV@gTypes@eB] (-1.9,1.1) -- (-2.2,1.4) node[label=(180+\cmdKV@general@rotate):\cmdKV@gLabels@eLB] {};
    \draw[\cmdKV@gTypes@eC] (-0.3,0.7) -- (0,0.7) node[label=(\cmdKV@general@rotate):\cmdKV@gLabels@eLC] {};
    \draw[\cmdKV@gTypes@eD] (-1.1,0.7) -- (-1.4,0.7) node[label=(180+\cmdKV@general@rotate):\cmdKV@gLabels@eLD] {};
    \draw[label distance=0,\cmdKV@gTypes@iA] (-1.9,0.3) -- node[label=(180+\cmdKV@general@rotate):\cmdKV@gLabels@iLA] {} (-1.9,1.1);
    \draw[label distance=0,\cmdKV@gTypes@iB] (-1.9,1.1) -- node[label=(90+\cmdKV@general@rotate):\cmdKV@gLabels@iLB] {} (-0.7,1.1);
    \draw[label distance=0,\cmdKV@gTypes@iC] (-0.7,1.1) -- node[label=(45+\cmdKV@general@rotate):\cmdKV@gLabels@iLC] {} (-0.3,0.7);
    \draw[label distance=0,\cmdKV@gTypes@iD] (-0.3,0.7) -- node[label=(-45+\cmdKV@general@rotate):\cmdKV@gLabels@iLD] {} (-0.7,0.3);
    \draw[label distance=0,\cmdKV@gTypes@iE] (-0.7,0.3) -- node[label=(-90+\cmdKV@general@rotate):\cmdKV@gLabels@iLE] {} (-1.9,0.3);
    \draw[label distance=0,\cmdKV@gTypes@iF] (-0.7,1.1) -- node[label=(180+\cmdKV@general@rotate):\cmdKV@gLabels@iLF] {} (-1.1,0.7);
    \draw[label distance=0,\cmdKV@gTypes@iG] (-1.1,0.7) -- node[label=(-180+\cmdKV@general@rotate):\cmdKV@gLabels@iLG] {} (-0.7,0.3);
    #2
  \end{tikzpicture}
  \gSetStdTypes{line}
}
\def\gTriPenta{\@ifnextchar[\@gTriPenta{\@gTriPenta[]}}
\def\@gTriPenta[#1]#2{%
  \setkeys*{pre}{#1}%
  \setrmkeys{general,gTypes,gLabels}%
  \begin{tikzpicture}
    [line width=1pt,
    baseline={([yshift=-0.5ex+\cmdKV@general@yshift]current bounding box.center)},
    scale=\cmdKV@general@scale,
    rotate=\cmdKV@general@rotate,
    font=\cmdKV@general@font]
    \draw[\cmdKV@gTypes@eA] (-108:0.5) -- (-135:1) node[label=(145+\cmdKV@general@rotate):\cmdKV@gLabels@eLA] {};
    \draw[\cmdKV@gTypes@eB] (180:0.5) -- (180:1) node[label=(180+\cmdKV@general@rotate):\cmdKV@gLabels@eLB] {};
    \draw[\cmdKV@gTypes@eC] (108:0.5) -- (135:1) node[label=(-145+\cmdKV@general@rotate):\cmdKV@gLabels@eLC] {};
    \draw[\cmdKV@gTypes@eD] (0:1.2) -- (0:1.5) node[label=(\cmdKV@general@rotate):\cmdKV@gLabels@eLD] {};
    \draw[label distance=0,\cmdKV@gTypes@iA] (-108:0.5) -- node[label=(-135+\cmdKV@general@rotate):\cmdKV@gLabels@iLA] {} (-180:0.5);
    \draw[label distance=0,\cmdKV@gTypes@iB] (180:0.5) -- node[label=(135+\cmdKV@general@rotate):\cmdKV@gLabels@iLB] {} (108:0.5);
    \draw[label distance=0,\cmdKV@gTypes@iC] (108:0.5) -- node[label=(72+\cmdKV@general@rotate):\cmdKV@gLabels@iLC] {} (36:0.5);
    \draw[label distance=0,\cmdKV@gTypes@iD] (36:0.5) -- node[label=(72+\cmdKV@general@rotate):\cmdKV@gLabels@iLD] {} (0:1.2);
    \draw[label distance=2,\cmdKV@gTypes@iE] (0:1.2) -- node[label=(-90+\cmdKV@general@rotate):\cmdKV@gLabels@iLE] {} (-36:0.5);
    \draw[label distance=2,\cmdKV@gTypes@iF] (-36:0.5) -- node[label=(-90+\cmdKV@general@rotate):\cmdKV@gLabels@iLF] {} (-108:0.5);
    \draw[label distance=0,\cmdKV@gTypes@iG] (36:0.5) -- node[label=(180+\cmdKV@general@rotate):\cmdKV@gLabels@iLG] {} (-36:0.5);
    #2
  \end{tikzpicture}
  \gSetStdTypes{line}
}
\def\gTriTri{\@ifnextchar[\@gTriTri{\@gTriTri[]}}
\def\@gTriTri[#1]#2{%
  \setkeys*{pre}{#1}%
  \setrmkeys{general,gTypes,gLabels}%
  \begin{tikzpicture}
    [line width=1pt,
    baseline={([yshift=-0.5ex+\cmdKV@general@yshift]current bounding box.center)},
    scale=\cmdKV@general@scale,
    rotate=\cmdKV@general@rotate,
    font=\cmdKV@general@font]
    \draw[\cmdKV@gTypes@eA] (0.3,0.3) -- (0,0) node[label=(-180+\cmdKV@general@rotate):\cmdKV@gLabels@eLA] {};
    \draw[\cmdKV@gTypes@eB] (0.3,1.1) -- (0,1.4) node[label=(180+\cmdKV@general@rotate):\cmdKV@gLabels@eLB] {};
    \draw[\cmdKV@gTypes@eC] (1.9,1.1) -- (2.2,1.4) node[label=(\cmdKV@general@rotate):\cmdKV@gLabels@eLC] {};
    \draw[\cmdKV@gTypes@eD] (1.9,0.3) -- (2.2,0) node[label=(\cmdKV@general@rotate):\cmdKV@gLabels@eLD] {};
    \draw[label distance=0,\cmdKV@gTypes@iA] (0.3,0.3) -- node[label=(180+\cmdKV@general@rotate):\cmdKV@gLabels@iLA] {} (0.3,1.1);
    \draw[label distance=0,\cmdKV@gTypes@iB] (0.3,1.1) -- node[label=(45+\cmdKV@general@rotate):\cmdKV@gLabels@iLB] {} (0.9,0.7);
    \draw[label distance=0,\cmdKV@gTypes@iC] (0.9,0.7) -- node[label=(-45+\cmdKV@general@rotate):\cmdKV@gLabels@iLC] {} (0.3,0.3);
    \draw[label distance=0,\cmdKV@gTypes@iD] (0.9,0.7) -- node[label=(90+\cmdKV@general@rotate):\cmdKV@gLabels@iLD] {} (1.3,0.7);
    \draw[label distance=0,\cmdKV@gTypes@iE] (1.3,0.7) -- node[label=(135+\cmdKV@general@rotate):\cmdKV@gLabels@iLE] {} (1.9,1.1);
    \draw[label distance=0,\cmdKV@gTypes@iF] (1.9,1.1) -- node[label=(\cmdKV@general@rotate):\cmdKV@gLabels@iLF] {} (1.9,0.3);
    \draw[label distance=0,\cmdKV@gTypes@iG] (1.9,0.3) -- node[label=(-135+\cmdKV@general@rotate):\cmdKV@gLabels@iLG] {} (1.3,0.7);
    #2
  \end{tikzpicture}
  \gSetStdTypes{line}
}

\def\gBoxBubA{\@ifnextchar[\@gBoxBubA{\@gBoxBubA[]}}
\def\@gBoxBubA[#1]#2{%
  \setkeys*{pre}{#1}%
  \setrmkeys{general,gTypes,gLabels}%
  \begin{tikzpicture}
    [line width=1pt,
    baseline={([yshift=-0.5ex+\cmdKV@general@yshift]current bounding box.center)},
    scale=\cmdKV@general@scale,
    rotate=\cmdKV@general@rotate,
    font=\cmdKV@general@font]
    \draw[\cmdKV@gTypes@eA] (0.1,0.1) -- (-0.1,-0.1) node[label=(-180+\cmdKV@general@rotate):\cmdKV@gLabels@eLA] {};
    \draw[\cmdKV@gTypes@eB] (0.1,0.9) -- (-0.1,1.1) node[label=(180+\cmdKV@general@rotate):\cmdKV@gLabels@eLB] {};
    \draw[\cmdKV@gTypes@eC] (0.9,0.9) -- (1.1,1.1) node[label=(\cmdKV@general@rotate):\cmdKV@gLabels@eLC] {};
    \draw[\cmdKV@gTypes@eD] (0.9,0.1) -- (1.1,-0.1) node[label=(\cmdKV@general@rotate):\cmdKV@gLabels@eLD] {};
    \draw[label distance=0,\cmdKV@gTypes@iA] (0.1,0.1) -- node[label=(180+\cmdKV@general@rotate):\cmdKV@gLabels@iLA] {} (0.1,0.9);
    \draw[label distance=0,\cmdKV@gTypes@iB] (0.1,0.9) -- node[label=(90+\cmdKV@general@rotate):\cmdKV@gLabels@iLB] {} (0.32,0.9);
    \draw[label distance=0,\cmdKV@gTypes@iC] (0.32,0.9) .. controls (0.32,1.15) and (0.68,1.15) .. node[label=(90+\cmdKV@general@rotate):\cmdKV@gLabels@iLC] {} (0.68,0.9);
    \draw[label distance=0,\cmdKV@gTypes@iD] (0.68,0.9) .. controls (0.68,0.65) and (0.32,0.65) .. node[label=(-90+\cmdKV@general@rotate):\cmdKV@gLabels@iLD] {} (0.32,0.9);
    \draw[label distance=0,\cmdKV@gTypes@iE] (0.68,0.9) -- node[label=(90+\cmdKV@general@rotate):\cmdKV@gLabels@iLE] {} (0.9,0.9);
    \draw[label distance=0,\cmdKV@gTypes@iF] (0.9,0.9) -- node[label=(\cmdKV@general@rotate):\cmdKV@gLabels@iLF] {} (0.9,0.1);
    \draw[label distance=0,\cmdKV@gTypes@iG] (0.9,0.1) -- node[label=(-90+\cmdKV@general@rotate):\cmdKV@gLabels@iLG] {} (0.1,0.1);
    #2
  \end{tikzpicture}
  \gSetStdTypes{line}
}

\def\gBoxBubB{\@ifnextchar[\@gBoxBubB{\@gBoxBubB[]}}
\def\@gBoxBubB[#1]#2{%
  \setkeys*{pre}{#1}%
  \setrmkeys{general,gTypes,gLabels}%
  \begin{tikzpicture}
    [line width=1pt,
    baseline={([yshift=-0.5ex+\cmdKV@general@yshift]current bounding box.center)},
    scale=\cmdKV@general@scale,
    rotate=\cmdKV@general@rotate,
    font=\cmdKV@general@font]
    \draw[label distance=-1,\cmdKV@gTypes@eA] (0,-0.4) -- (0,-0.7) node[label=(180+\cmdKV@general@rotate):\cmdKV@gLabels@eLA] {};
    \draw[\cmdKV@gTypes@eB] (-0.4,0) -- (-0.7,0) node[label=(180+\cmdKV@general@rotate):\cmdKV@gLabels@eLB] {};
    \draw[label distance=-1,\cmdKV@gTypes@eC] (0,0.4) -- (0,0.7) node[label=(180+\cmdKV@general@rotate):\cmdKV@gLabels@eLC] {};
    \draw[\cmdKV@gTypes@eD] (1.1,0) -- (1.4,0) node[label=(\cmdKV@general@rotate):\cmdKV@gLabels@eLD] {};
    \draw[label distance=0,\cmdKV@gTypes@iA] (0,-0.4) -- node[label=(-135+\cmdKV@general@rotate):\cmdKV@gLabels@iLA] {} (-0.4,0);
    \draw[label distance=0,\cmdKV@gTypes@iB] (-0.4,0) -- node[label=(135+\cmdKV@general@rotate):\cmdKV@gLabels@iLB] {} (0,0.4);
    \draw[label distance=0,\cmdKV@gTypes@iC] (0,0.4) -- node[label=(45+\cmdKV@general@rotate):\cmdKV@gLabels@iLC] {} (0.4,0);
    \draw[label distance=0,\cmdKV@gTypes@iD] (0.4,0) -- node[label=(-45+\cmdKV@general@rotate):\cmdKV@gLabels@iLD] {} (0,-0.4);
    \draw[label distance=0,\cmdKV@gTypes@iE] (0.4,0) --  node[label=(90+\cmdKV@general@rotate):\cmdKV@gLabels@iLE] {} (0.7,0);
    \draw[label distance=0,\cmdKV@gTypes@iF] (0.7,0) .. controls (0.7,0.26) and (1.1,0.26) .. node[label=(90+\cmdKV@general@rotate):\cmdKV@gLabels@iLF] {} (1.1,0);
    \draw[label distance=0,\cmdKV@gTypes@iG] (1.1,0) .. controls (1.1,-0.26) and (0.7,-0.26) ..  node[label=(-90+\cmdKV@general@rotate):\cmdKV@gLabels@iLG] {} (0.7,0);
    #2
  \end{tikzpicture}
  \gSetStdTypes{line}
}

\def\gPentaTad{\@ifnextchar[\@gPentaTad{\@gPentaTad[]}}
\def\@gPentaTad[#1]#2{%
  \setkeys*{pre}{#1}%
  \setrmkeys{general,gTypes,gLabels}%
  \begin{tikzpicture}
    [line width=1pt,
    baseline={([yshift=-0.5ex+\cmdKV@general@yshift]current bounding box.center)},
    scale=\cmdKV@general@scale,
    rotate=\cmdKV@general@rotate,
    font=\cmdKV@general@font]
    \draw[label distance=-0.5,\cmdKV@gTypes@eA] (108:0.35) -- (108:0.75) node[label=(180+\cmdKV@general@rotate):\cmdKV@gLabels@eLA] {};
    \draw[\cmdKV@gTypes@eB] (36:0.35) -- (36:0.75) node[label=(36+\cmdKV@general@rotate):\cmdKV@gLabels@eLB] {};
    \draw[\cmdKV@gTypes@eC] (-36:0.35) -- (-36:0.75) node[label=(-36+\cmdKV@general@rotate):\cmdKV@gLabels@eLC] {};
    \draw[label distance=-0.5,\cmdKV@gTypes@eD] (-108:0.35) -- (-108:0.75) node[label=(-180+\cmdKV@general@rotate):\cmdKV@gLabels@eLD] {};
    \draw[label distance=0,\cmdKV@gTypes@iA] (108:0.35) -- node[label=(72+\cmdKV@general@rotate):\cmdKV@gLabels@iLA] {} (36:0.35);
    \draw[label distance=0,\cmdKV@gTypes@iB] (36:0.35) -- node[label=(\cmdKV@general@rotate):\cmdKV@gLabels@iLB] {} (-36:0.35);
    \draw[label distance=0,\cmdKV@gTypes@iC] (-36:0.35) -- node[label=(-72+\cmdKV@general@rotate):\cmdKV@gLabels@iLC] {} (-108:0.35);
    \draw[label distance=0,\cmdKV@gTypes@iD] (-108:0.35) -- node[label=(-144+\cmdKV@general@rotate):\cmdKV@gLabels@iLD] {} (-180:0.35);
    \draw[label distance=0,\cmdKV@gTypes@iE] (180:0.35) -- node[label=(144+\cmdKV@general@rotate):\cmdKV@gLabels@iLE] {} (108:0.35);
    \draw[label distance=0,\cmdKV@gTypes@iF] (180:0.35) -- node[label=(90+\cmdKV@general@rotate):\cmdKV@gLabels@iLF] {} (180:0.75);
    \draw[label distance=-0.5,\cmdKV@gTypes@iG] (180:0.75) .. controls ($(180:0.75)+(0,-0.18)$) and ($(180:1)+(0,-0.18)$) .. (180:1) node[label=(180+\cmdKV@general@rotate):\cmdKV@gLabels@iLG] {} .. controls ($(180:1)+(0,0.18)$) and ($(180:0.75)+(0,0.18)$) .. (180:0.75);
    #2
  \end{tikzpicture}
  \gSetStdTypes{line}
}

\def\gBoxTadA{\@ifnextchar[\@gBoxTadA{\@gBoxTadA[]}}
\def\@gBoxTadA[#1]#2{%
  \setkeys*{pre}{#1}%
  \setrmkeys{general,gTypes,gLabels}%
  \begin{tikzpicture}
    [line width=1pt,
    baseline={([yshift=-0.5ex+\cmdKV@general@yshift]current bounding box.center)},
    scale=\cmdKV@general@scale,
    rotate=\cmdKV@general@rotate,
    font=\cmdKV@general@font]
    \draw[label distance=-0.5,\cmdKV@gTypes@eA] (0,0.4) -- (0,0.7) node[label=(\cmdKV@general@rotate):\cmdKV@gLabels@eLA] {};
    \draw[\cmdKV@gTypes@eB] (0.4,0) -- (0.7,0) node[label=(\cmdKV@general@rotate):\cmdKV@gLabels@eLB] {};
    \draw[label distance=-0.5,\cmdKV@gTypes@eC] (0,-0.4) -- (0,-0.7) node[label=(\cmdKV@general@rotate):\cmdKV@gLabels@eLC] {};
    \draw[\cmdKV@gTypes@eD] (-0.7,0) -- (-0.7,-0.3) node[label=(-90+\cmdKV@general@rotate):\cmdKV@gLabels@eLD] {};
    \draw[label distance=0,\cmdKV@gTypes@iA] (0,0.4) -- node[label=(45+\cmdKV@general@rotate):\cmdKV@gLabels@iLA] {} (0.4,0);
    \draw[label distance=0,\cmdKV@gTypes@iB] (0.4,0) -- node[label=(-45+\cmdKV@general@rotate):\cmdKV@gLabels@iLB] {} (0,-0.4);
    \draw[label distance=0,\cmdKV@gTypes@iC] (0,-0.4) -- node[label=(-135+\cmdKV@general@rotate):\cmdKV@gLabels@iLC] {} (-0.4,0);
    \draw[label distance=0,\cmdKV@gTypes@iD] (-0.4,0) -- node[label=(135+\cmdKV@general@rotate):\cmdKV@gLabels@iLD] {} (0,0.4);
    \draw[label distance=0,\cmdKV@gTypes@iE] (-0.4,0) --  node[label=(90+\cmdKV@general@rotate):\cmdKV@gLabels@iLE] {} (-0.7,0);
    \draw[label distance=0,\cmdKV@gTypes@iF] (-0.7,0) -- node[label=(90+\cmdKV@general@rotate):\cmdKV@gLabels@iLF] {} (-1.1,0);
    \draw[label distance=-0.5,\cmdKV@gTypes@iG] (-1.1,0) .. controls (-1.1,-0.18) and (-1.35,-0.18) .. (-1.35,0) node[label=(180+\cmdKV@general@rotate):\cmdKV@gLabels@iLG] {} .. controls (-1.35,0.18) and (-1.1,0.18) .. (-1.1,0);
    #2
  \end{tikzpicture}
  \gSetStdTypes{line}
}

\def\gBoxBoxBoxA{\@ifnextchar[\@gBoxBoxBoxA{\@gBoxBoxBoxA[]}}
\def\@gBoxBoxBoxA[#1]#2{%
  \setkeys*{pre}{#1}%
  \setrmkeys{general,gTypes,gLabels}%
  \begin{tikzpicture}
    [line width=1pt,
    baseline={([yshift=-0.5ex+\cmdKV@general@yshift]current bounding box.center)},
    scale=\cmdKV@general@scale,
    rotate=\cmdKV@general@rotate,
    font=\cmdKV@general@font]
    \draw[\cmdKV@gTypes@eA] (-0.3,0.3) -- (-0.6,0) node[label=(-180+\cmdKV@general@rotate):\cmdKV@gLabels@eLA] {};
    \draw[\cmdKV@gTypes@eB] (-0.3,0.9) -- (-0.6,1.2) node[label=(180+\cmdKV@general@rotate):\cmdKV@gLabels@eLB] {};
    \draw[\cmdKV@gTypes@eC] (1.5,0.9) -- (1.8,1.2) node[label=(\cmdKV@general@rotate):\cmdKV@gLabels@eLC] {};
    \draw[\cmdKV@gTypes@eD] (1.5,0.3) -- (1.8,0) node[label=(\cmdKV@general@rotate):\cmdKV@gLabels@eLD] {};
    \draw[label distance=0,\cmdKV@gTypes@iA] (-0.3,0.3) -- node[label=(180+\cmdKV@general@rotate):\cmdKV@gLabels@iLA] {} (-0.3,0.9);
    \draw[label distance=0,\cmdKV@gTypes@iB] (-0.3,0.9) -- node[label=(90+\cmdKV@general@rotate):\cmdKV@gLabels@iLB] {} (0.3,0.9);
    \draw[label distance=0,\cmdKV@gTypes@iC] (0.3,0.9) -- node[label=(90+\cmdKV@general@rotate):\cmdKV@gLabels@iLC] {} (0.9,0.9);
    \draw[label distance=0,\cmdKV@gTypes@iD] (0.9,0.9) -- node[label=(90+\cmdKV@general@rotate):\cmdKV@gLabels@iLD] {} (1.5,0.9);
    \draw[label distance=0,\cmdKV@gTypes@iE] (1.5,0.9) -- node[label=(\cmdKV@general@rotate):\cmdKV@gLabels@iLE] {} (1.5,0.3);
    \draw[label distance=0,\cmdKV@gTypes@iF] (1.5,0.3) -- node[label=(-90+\cmdKV@general@rotate):\cmdKV@gLabels@iLF] {} (0.9,0.3);
    \draw[label distance=0,\cmdKV@gTypes@iG] (0.9,0.3) -- node[label=(-90+\cmdKV@general@rotate):\cmdKV@gLabels@iLG] {} (0.3,0.3);
    \draw[label distance=0,\cmdKV@gTypes@iH] (0.3,0.3) -- node[label=(-90+\cmdKV@general@rotate):\cmdKV@gLabels@iLH] {} (-0.3,0.3);
    \draw[label distance=0,\cmdKV@gTypes@iI] (0.3,0.9) -- node[label=(\cmdKV@general@rotate):\cmdKV@gLabels@iLI] {} (0.3,0.3);
    \draw[label distance=0,\cmdKV@gTypes@iJ] (0.9,0.9) -- node[label=(\cmdKV@general@rotate):\cmdKV@gLabels@iLJ] {} (0.9,0.3);
    #2
  \end{tikzpicture}
  \gSetStdTypes{line}
}

\def\gBoxBoxBoxB{\@ifnextchar[\@gBoxBoxBoxB{\@gBoxBoxBoxB[]}}
\def\@gBoxBoxBoxB[#1]#2{%
  \setkeys*{pre}{#1}%
  \setrmkeys{general,gTypes,gLabels}%
  \begin{tikzpicture}
    [line width=1pt,
    baseline={([yshift=-0.5ex+\cmdKV@general@yshift]current bounding box.center)},
    scale=\cmdKV@general@scale,
    rotate=\cmdKV@general@rotate,
    font=\cmdKV@general@font]
    \draw[\cmdKV@gTypes@eA] (0.3,0.3) -- (0,0) node[label=(-180+\cmdKV@general@rotate):\cmdKV@gLabels@eLA] {};
    \draw[\cmdKV@gTypes@eB] (0.3,1.5) -- (0,1.8) node[label=(180+\cmdKV@general@rotate):\cmdKV@gLabels@eLB] {};
    \draw[\cmdKV@gTypes@eC] (1.5,1.5) -- (1.8,1.8) node[label=(\cmdKV@general@rotate):\cmdKV@gLabels@eLC] {};
    \draw[\cmdKV@gTypes@eD] (1.5,0.3) -- (1.8,0) node[label=(\cmdKV@general@rotate):\cmdKV@gLabels@eLD] {};
    \draw[label distance=0,\cmdKV@gTypes@iA] (0.3,0.3) -- node[label=(180+\cmdKV@general@rotate):\cmdKV@gLabels@iLA] {} (0.3,0.9);
    \draw[label distance=0,\cmdKV@gTypes@iB] (0.3,0.9) -- node[label=(180+\cmdKV@general@rotate):\cmdKV@gLabels@iLB] {} (0.3,1.5);
    \draw[label distance=0,\cmdKV@gTypes@iC] (0.3,1.5) -- node[label=(90+\cmdKV@general@rotate):\cmdKV@gLabels@iLC] {} (0.9,1.5);
    \draw[label distance=0,\cmdKV@gTypes@iD] (0.9,1.5) -- node[label=(90+\cmdKV@general@rotate):\cmdKV@gLabels@iLD] {} (1.5,1.5);
    \draw[label distance=0,\cmdKV@gTypes@iE] (1.5,1.5) -- node[label=(\cmdKV@general@rotate):\cmdKV@gLabels@iLE] {} (1.5,0.3);
    \draw[label distance=0,\cmdKV@gTypes@iF] (1.5,0.3) -- node[label=(-90+\cmdKV@general@rotate):\cmdKV@gLabels@iLF] {} (0.9,0.3);
    \draw[label distance=0,\cmdKV@gTypes@iG] (0.9,0.3) -- node[label=(-90+\cmdKV@general@rotate):\cmdKV@gLabels@iLG] {} (0.3,0.3);
    \draw[label distance=0,\cmdKV@gTypes@iH] (0.9,1.5) -- node[label=(\cmdKV@general@rotate):\cmdKV@gLabels@iLH] {} (0.9,0.9);
    \draw[label distance=0,\cmdKV@gTypes@iI] (0.9,0.9) -- node[label=(\cmdKV@general@rotate):\cmdKV@gLabels@iLI] {} (0.9,0.3);
    \draw[label distance=0,\cmdKV@gTypes@iJ] (0.9,0.9) -- node[label=(90+\cmdKV@general@rotate):\cmdKV@gLabels@iLJ] {} (0.3,0.9);
    #2
  \end{tikzpicture}
  \gSetStdTypes{line}
}

\def\cTree{\@ifnextchar[\@cTree{\@cTree[]}}
\def\@cTree[#1]#2{%
  \setkeys*{pre}{#1}%
  \setrmkeys{general,gTypes,gLabels,blobs}%
  \begin{tikzpicture}
    [line width=1pt,
    baseline={([yshift=-0.5ex+\cmdKV@general@yshift]current bounding box.center)},
    scale=\cmdKV@general@scale,
    rotate=\cmdKV@general@rotate,
    font=\cmdKV@general@font]
    \fill[\cmdKV@blobs@blobA] (0,0) ellipse (0.25 and 0.25);
    \draw[\cmdKV@gTypes@eA] ($ (0,0) + (-135:0.25) $) -- ($ (0,0) + (-135:0.6) $) node[label=(-180+\cmdKV@general@rotate):\cmdKV@gLabels@eLA] {};
    \draw[\cmdKV@gTypes@eB] ($ (0,0) + (135:0.25) $) -- ($ (0,0) + (135:0.6) $) node[label=(180+\cmdKV@general@rotate):\cmdKV@gLabels@eLB] {};
    \draw[\cmdKV@gTypes@eC] ($ (0,0) + (45:0.25) $) -- ($ (0,0) + (45:0.6) $) node[label=(\cmdKV@general@rotate):\cmdKV@gLabels@eLC] {};
    \draw[\cmdKV@gTypes@eD] ($ (0,0) + (-45:0.25) $) -- ($ (0,0) + (-45:0.6) $) node[label=(\cmdKV@general@rotate):\cmdKV@gLabels@eLD] {};
    \node[label=(90+\cmdKV@general@rotate):\cmdKV@gLabels@iLA] at (0,0.3) {};
    \node[label=(\cmdKV@general@rotate):\cmdKV@gLabels@iLB] at (0.3,0) {};
    #2
  \end{tikzpicture}
  \gSetStdTypes{line}
}

\def\cBoxA{\@ifnextchar[\@cBoxA{\@cBoxA[]}}
\def\@cBoxA[#1]#2{%
  \setkeys*{pre}{#1}%
  \setrmkeys{general,gTypes,gLabels,dots,blobs}%
  \begin{tikzpicture}
    [line width=1pt,
    baseline={([yshift=-0.5ex+\cmdKV@general@yshift]current bounding box.center)},
    scale=\cmdKV@general@scale,
    rotate=\cmdKV@general@rotate,
    font=\cmdKV@general@font]
    \fill[\cmdKV@blobs@blobA] (-0.4,0) ellipse (0.2 and 0.4);
    \fill[\cmdKV@blobs@blobB] (0.4,0) ellipse (0.2 and 0.4);
    \draw[\cmdKV@gTypes@eA] ($ (-0.4,0) + (-135:0.26) $) -- ($ (-0.4,0) + (-135:0.7) $) node[label=(-180+\cmdKV@general@rotate):\cmdKV@gLabels@eLA] {};
    \draw[\cmdKV@gTypes@eB] ($ (-0.4,0) + (135:0.26) $) -- ($ (-0.4,0) + (135:0.7) $) node[label=(180+\cmdKV@general@rotate):\cmdKV@gLabels@eLB] {};
    \draw[\cmdKV@gTypes@eC] ($ (0.4,0) + (45:0.26) $) -- ($ (0.4,0) + (45:0.7) $) node[label=(\cmdKV@general@rotate):\cmdKV@gLabels@eLC] {};
    \draw[\cmdKV@gTypes@eD] ($ (0.4,0) + (-45:0.26) $) -- ($ (0.4,0) + (-45:0.7) $) node[label=(\cmdKV@general@rotate):\cmdKV@gLabels@eLD] {};
    \draw[label distance=1,\cmdKV@gTypes@iA] ($ (-0.4,0) + (45:0.26) $) -- node[label=(90+\cmdKV@general@rotate):\cmdKV@gLabels@iLA] {}($ (0.4,0) + (135:0.26) $);
    \draw[label distance=1,\cmdKV@gTypes@iB] ($ (0.4,0) + (-135:0.26) $) -- node[label=(-90+\cmdKV@general@rotate):\cmdKV@gLabels@iLB] {}($ (-0.4,0) + (-45:0.26) $);
    \node[label=(180+\cmdKV@general@rotate):\cmdKV@gLabels@iLC] at (-0.6,0) {};
    \node[label=(\cmdKV@general@rotate):\cmdKV@gLabels@iLD] at (0.6,0) {};
    \ifKV@dots@lDots
      \draw[dotted] ($(-0.4,0) + (-150:0.7)$) to[bend left] ($(-0.4,0) + (150:0.7)$);%
    \fi
    \ifKV@dots@rDots
      \draw[dotted] ($(0.4,0) + (30:0.7)$) to[bend left] ($(0.4,0) + (-30:0.7)$);%
    \fi
    #2
  \end{tikzpicture}
  \gSetStdTypes{line}
}

\def\cBoxAA{\@ifnextchar[\@cBoxAA{\@cBoxAA[]}}
\def\@cBoxAA[#1]#2{%
  \setkeys*{pre}{#1}%
  \setrmkeys{general,gTypes,gLabels,dots,blobs}%
  \begin{tikzpicture}
    [line width=1pt,
    baseline={([yshift=-0.5ex+\cmdKV@general@yshift]current bounding box.center)},
    scale=\cmdKV@general@scale,
    rotate=\cmdKV@general@rotate,
    font=\cmdKV@general@font]
    \fill[\cmdKV@blobs@blobA] (-0.4,0) ellipse (0.2 and 0.4);
    \fill[\cmdKV@blobs@blobB] (0.4,0) ellipse (0.2 and 0.4);
    \draw[\cmdKV@gTypes@eA] ($ (-0.4,0) + (-135:0.26) $) -- ($ (-0.4,0) + (-135:0.7) $) node[label=(-180+\cmdKV@general@rotate):\cmdKV@gLabels@eLA] {};
    \draw[\cmdKV@gTypes@eB] ($ (-0.4,0) + (135:0.26) $) -- ($ (-0.4,0) + (135:0.7) $) node[label=(180+\cmdKV@general@rotate):\cmdKV@gLabels@eLB] {};
    \draw[\cmdKV@gTypes@eC] ($ (0.4,0) + (45:0.26) $) -- ($ (0.4,0) + (45:0.7) $) node[label=(\cmdKV@general@rotate):\cmdKV@gLabels@eLC] {};
    \draw[\cmdKV@gTypes@eD] ($ (0.4,0) + (-45:0.26) $) -- ($ (0.4,0) + (-45:0.7) $) node[label=(\cmdKV@general@rotate):\cmdKV@gLabels@eLD] {};
    \draw[label distance=1,\cmdKV@gTypes@iA] ($ (-0.4,0) + (60:0.32) $) to[bend left] node[label=(90+\cmdKV@general@rotate):\cmdKV@gLabels@iLA] {}($ (0.4,0) + (120:0.32) $);
    \draw[label distance=1,\cmdKV@gTypes@iB] ($ (0.4,0) + (-120:0.32) $) to[bend left]  node[label=(-90+\cmdKV@general@rotate):\cmdKV@gLabels@iLB] {}($ (-0.4,0) + (-60:0.32) $);
    \draw[dotted] (0,0.25) -- (0,-0.25);
    \node[label=(180+\cmdKV@general@rotate):\cmdKV@gLabels@iLC] at (-0.6,0) {};
    \node[label=(\cmdKV@general@rotate):\cmdKV@gLabels@iLD] at (0.6,0) {};
    \ifKV@dots@lDots
      \draw[dotted] ($(-0.4,0) + (-150:0.7)$) to[bend left] ($(-0.4,0) + (150:0.7)$);%
    \fi
    \ifKV@dots@rDots
      \draw[dotted] ($(0.4,0) + (30:0.7)$) to[bend left] ($(0.4,0) + (-30:0.7)$);%
    \fi
    #2
  \end{tikzpicture}
  \gSetStdTypes{line}
}

\def\cBoxB{\@ifnextchar[\@cBoxB{\@cBoxB[]}}
\def\@cBoxB[#1]#2{%
  \setkeys*{pre}{#1}%
  \setrmkeys{general,gTypes,gLabels,blobs}%
  \begin{tikzpicture}
    [line width=1pt,
    baseline={([yshift=-0.5ex+\cmdKV@general@yshift]current bounding box.center)},
    scale=\cmdKV@general@scale,
    rotate=\cmdKV@general@rotate,
    font=\cmdKV@general@font]
    \fill[\cmdKV@blobs@blobA] (0,-0.3) ellipse (0.4 and 0.2);
    \fill[\cmdKV@blobs@blobB] (0,0.3) ellipse (0.4 and 0.2);
    \draw[\cmdKV@gTypes@eA] ($ (0,-0.3) + (180:0.4) $) -- ($ (0,-0.3) + (-170:0.7) $) node[label=(180+\cmdKV@general@rotate):\cmdKV@gLabels@eLA] {};
    \draw[\cmdKV@gTypes@eB] ($ (0,0.3) + (180:0.4) $) -- ($ (0,0.3) + (170:0.7) $) node[label=(180+\cmdKV@general@rotate):\cmdKV@gLabels@eLB] {};
    \draw[\cmdKV@gTypes@eC] ($ (0,0.3) + (0:0.4) $) -- ($ (0,0.3) + (10:0.7) $) node[label=(\cmdKV@general@rotate):\cmdKV@gLabels@eLC] {};
    \draw[\cmdKV@gTypes@eD] ($ (0,-0.3) + (0:0.4) $) -- ($ (0,-0.3) + (-10:0.7) $) node[label=(\cmdKV@general@rotate):\cmdKV@gLabels@eLD] {};
    \draw[label distance=1,\cmdKV@gTypes@iA] ($ (0,-0.3) + (135:0.26) $) -- node[label=(-180+\cmdKV@general@rotate):\cmdKV@gLabels@iLA] {}($ (0,0.3) + (-135:0.26) $);
    \draw[label distance=1,\cmdKV@gTypes@iB] ($ (0,0.3) + (-45:0.26) $) -- node[label=(\cmdKV@general@rotate):\cmdKV@gLabels@iLB] {}($ (0,-0.3) + (45:0.26) $);
    \node[label=(90+\cmdKV@general@rotate):\cmdKV@gLabels@iLC] at (0,0.5) {};
    \node[label=(-90+\cmdKV@general@rotate):\cmdKV@gLabels@iLD] at (0,-0.5) {};
    #2
  \end{tikzpicture}
  \gSetStdTypes{line}
}

\def\cBoxBoxA{\@ifnextchar[\@cBoxBoxA{\@cBoxBoxA[]}}
\def\@cBoxBoxA[#1]#2{%
  \setkeys*{pre}{#1}%
  \setrmkeys{general,gTypes,gLabels,blobs}%
  \begin{tikzpicture}
    [line width=1pt,
    baseline={([yshift=-0.5ex+\cmdKV@general@yshift]current bounding box.center)},
    scale=\cmdKV@general@scale,
    rotate=\cmdKV@general@rotate,
    font=\cmdKV@general@font]
    \fill[\cmdKV@blobs@blobA] (-0.4,0) ellipse (0.2 and 0.4);
    \fill[\cmdKV@blobs@blobB] (0.4,0) ellipse (0.2 and 0.4);
    \fill[\cmdKV@blobs@blobC] (1.2,0) ellipse (0.2 and 0.4);
    \draw[\cmdKV@gTypes@eA] ($ (-0.4,0) + (-135:0.26) $) -- ($ (-0.4,0) + (-135:0.7) $) node[label=(-180+\cmdKV@general@rotate):\cmdKV@gLabels@eLA] {};
    \draw[\cmdKV@gTypes@eB] ($ (-0.4,0) + (135:0.26) $) -- ($ (-0.4,0) + (135:0.7) $) node[label=(180+\cmdKV@general@rotate):\cmdKV@gLabels@eLB] {};
    \draw[\cmdKV@gTypes@eC] ($ (1.2,0) + (45:0.26) $) -- ($ (1.2,0) + (45:0.7) $) node[label=(\cmdKV@general@rotate):\cmdKV@gLabels@eLC] {};
    \draw[\cmdKV@gTypes@eD] ($ (1.2,0) + (-45:0.26) $) -- ($ (1.2,0) + (-45:0.7) $) node[label=(\cmdKV@general@rotate):\cmdKV@gLabels@eLD] {};
    \draw[label distance=1,\cmdKV@gTypes@iA] ($ (-0.4,0) + (45:0.26) $) -- node[label=(90+\cmdKV@general@rotate):\cmdKV@gLabels@iLA] {}($ (0.4,0) + (135:0.26) $);
    \draw[label distance=1,\cmdKV@gTypes@iB] ($ (0.4,0) + (45:0.26) $) -- node[label=(90+\cmdKV@general@rotate):\cmdKV@gLabels@iLB] {}($ (1.2,0) + (135:0.26) $);
    \draw[label distance=1,\cmdKV@gTypes@iC] ($ (1.2,0) + (-135:0.26) $) -- node[label=(-90+\cmdKV@general@rotate):\cmdKV@gLabels@iLC] {}($ (0.4,0) + (-45:0.26) $);
    \draw[label distance=1,\cmdKV@gTypes@iD] ($ (0.4,0) + (-135:0.26) $) -- node[label=(-90+\cmdKV@general@rotate):\cmdKV@gLabels@iLD] {}($ (-0.4,0) + (-45:0.26) $);
    \node[label=(180+\cmdKV@general@rotate):\cmdKV@gLabels@iLE] at (-0.6,0) {};
    \node[label=(\cmdKV@general@rotate):\cmdKV@gLabels@iLF] at (0.6,0) {};
    \node[label=(\cmdKV@general@rotate):\cmdKV@gLabels@iLG] at (1.4,0) {};
    #2
  \end{tikzpicture}
  \gSetStdTypes{line}
}

\def\cBoxBoxB{\@ifnextchar[\@cBoxBoxB{\@cBoxBoxB[]}}
\def\@cBoxBoxB[#1]#2{%
  \setkeys*{pre}{#1}%
  \setrmkeys{general,gTypes,gLabels}%
  \begin{tikzpicture}
    [line width=1pt,
    baseline={([yshift=-0.5ex+\cmdKV@general@yshift]current bounding box.center)},
    scale=\cmdKV@general@scale,
    rotate=\cmdKV@general@rotate,
    font=\cmdKV@general@font]
    \fill[\cmdKV@blobs@blobA] (0,0.3) ellipse (0.4 and 0.2);
    \fill[\cmdKV@blobs@blobB] (0,-0.3) ellipse (0.4 and 0.2);
    \fill[\cmdKV@blobs@blobC] (1,0) ellipse (0.2 and 0.4);
    \draw[\cmdKV@gTypes@eA] ($ (0,-0.3) + (180:0.4) $) -- ($ (0,-0.3) + (-170:0.7) $) node[label=(180+\cmdKV@general@rotate):\cmdKV@gLabels@eLA] {};
    \draw[\cmdKV@gTypes@eB] ($ (0,0.3) + (180:0.4) $) -- ($ (0,0.3) + (170:0.7) $) node[label=(180+\cmdKV@general@rotate):\cmdKV@gLabels@eLB] {};
    \draw[\cmdKV@gTypes@eC] ($ (1,0) + (45:0.26) $) -- ($ (1,0) + (45:0.7) $) node[label=(\cmdKV@general@rotate):\cmdKV@gLabels@eLC] {};
    \draw[\cmdKV@gTypes@eD] ($ (1,0) + (-45:0.26) $) -- ($ (1,0) + (-45:0.7) $) node[label=(\cmdKV@general@rotate):\cmdKV@gLabels@eLD] {};
    \draw[label distance=1,\cmdKV@gTypes@iA] ($ (0,-0.3) + (135:0.26) $) -- node[label=(-180+\cmdKV@general@rotate):\cmdKV@gLabels@iLA] {}($ (0,0.3) + (-135:0.26) $);
    \draw[label distance=1,\cmdKV@gTypes@iB] ($ (0,0.3) + (0:0.4) $) -- node[label=(90+\cmdKV@general@rotate):\cmdKV@gLabels@iLB] {} ($ (1,0) + (135:0.26) $);
    \draw[label distance=1,\cmdKV@gTypes@iC] ($ (1,0) + (-135:0.26) $) -- node[label=(-90+\cmdKV@general@rotate):\cmdKV@gLabels@iLC] {} ($ (0,-0.3) + (0:0.4) $);
    \draw[label distance=1,\cmdKV@gTypes@iD] ($ (0,0.3) + (-45:0.26) $) -- node[label=(\cmdKV@general@rotate):\cmdKV@gLabels@iLD] {}($ (0,-0.3) + (45:0.26) $);
    \node[label=(90+\cmdKV@general@rotate):\cmdKV@gLabels@iLE] at (0,0.5) {};
    \node[label=(\cmdKV@general@rotate):\cmdKV@gLabels@iLF] at (1.2,0) {};
    \node[label=(-90+\cmdKV@general@rotate):\cmdKV@gLabels@iLG] at (0,-0.5) {};
    #2
  \end{tikzpicture}
  \gSetStdTypes{line}
}

\def\cNPBox{\@ifnextchar[\@cNPBox{\@cNPBox[]}}
\def\@cNPBox[#1]#2{%
  \setkeys*{pre}{#1}%
  \setrmkeys{general,gTypes,gLabels,blob}%
  \begin{tikzpicture}
    [line width=1pt,
    baseline={([yshift=-0.5ex+\cmdKV@general@yshift]current bounding box.center)},
    scale=\cmdKV@general@scale,
    rotate=\cmdKV@general@rotate,
    font=\cmdKV@general@font]
    \fill[\cmdKV@blobs@blobA] (-0.6,0) ellipse (0.2 and 0.4);
    \fill[\cmdKV@blobs@blobB] (0.4,0) ellipse (0.2 and 0.4);
    \draw[\cmdKV@gTypes@eA] ($ (-0.6,0) + (-180:0.2) $) -- ($ (-0.6,0) + (-180:0.6) $) node[label=(-180+\cmdKV@general@rotate):\cmdKV@gLabels@eLA] {};
    \draw[\cmdKV@gTypes@eB] ($ (-0.6,0) + (0:0.2) $) -- (-0.18,0) node[label=(\cmdKV@general@rotate):\cmdKV@gLabels@eLB] {};
    \draw[\cmdKV@gTypes@eC] ($ (0.4,0) + (45:0.26) $) -- ($ (0.4,0) + (45:0.7) $) node[label=(\cmdKV@general@rotate):\cmdKV@gLabels@eLC] {};
    \draw[\cmdKV@gTypes@eD] ($ (0.4,0) + (-45:0.26) $) -- ($ (0.4,0) + (-45:0.7) $) node[label=(\cmdKV@general@rotate):\cmdKV@gLabels@eLD] {};
    \draw[label distance=1,\cmdKV@gTypes@iA] ($ (-0.6,0) + (60:0.3) $) -- node[label=(90+\cmdKV@general@rotate):\cmdKV@gLabels@iLA] {}($ (0.4,0) + (135:0.26) $);
    \draw[label distance=1,\cmdKV@gTypes@iB] ($ (0.4,0) + (-135:0.26) $) -- node[label=(-90+\cmdKV@general@rotate):\cmdKV@gLabels@iLB] {}($ (-0.6,0) + (-60:0.3) $);
    \node[label=(\cmdKV@general@rotate):\cmdKV@gLabels@iLD] at (0.6,0) {};
    #2
  \end{tikzpicture}
  \gSetStdTypes{line}
}

\def\cFourPtRec{\@ifnextchar[\@cFourPtRec{\@cFourPtRec[]}}
\def\@cFourPtRec[#1]#2{%
  \setkeys*{pre}{#1}%
  \setrmkeys{general,gTypes,gLabels,blobs}%
  \begin{tikzpicture}
    [line width=1pt,
    baseline={([yshift=-0.5ex+\cmdKV@general@yshift]current bounding box.center)},
    scale=\cmdKV@general@scale,
    rotate=\cmdKV@general@rotate,
    font=\cmdKV@general@font]
    \fill[\cmdKV@blobs@blobA] (-0.5,0) ellipse (0.2 and 0.3);
    \fill[\cmdKV@blobs@blobB] (0.5,0) ellipse (0.3 and 0.4);
    \filldraw[fill=white,draw=black] (0.55,0.15) circle (0.1);
    \filldraw[fill=white,draw=black] (0.45,-0.1) circle (0.1);
    \draw[\cmdKV@gTypes@eA] ($(-0.5,0) + (135:0.23)$) -- ($(-0.5,0) + (135:0.5)$) node[label=(180+\cmdKV@general@rotate):\cmdKV@gLabels@eLA] {};
    \draw[\cmdKV@gTypes@eB] ($(-0.5,0) + (-135:0.23)$) -- ($(-0.5,0) + (-135:0.5)$) node[label=(180+\cmdKV@general@rotate):\cmdKV@gLabels@eLB] {};
    \draw[\cmdKV@gTypes@eC] ($(0.5,0) + (45:0.35)$) -- ($(0.5,0) + (45:0.7)$) node[label=(\cmdKV@general@rotate):\cmdKV@gLabels@eLC] {};
    \draw[\cmdKV@gTypes@eD] ($(0.5,0) + (-45:0.35)$) -- ($(0.5,0) + (-45:0.7)$) node[label=(\cmdKV@general@rotate):\cmdKV@gLabels@eLD] {};
    \draw[label distance=1,\cmdKV@gTypes@iA] ($(-0.5,0) + (30:0.23)$) -- node[label=(90+\cmdKV@general@rotate):\cmdKV@gLabels@iLA] {} ($(0.5,0) + (157:0.3)$);
    \draw[label distance=1,\cmdKV@gTypes@iB] ($(0.5,0) + (-157:0.3)$) -- node[label=(-90+\cmdKV@general@rotate):\cmdKV@gLabels@iLB] {} ($(-0.5,0) + (-30:0.23)$);
    #2
 \end{tikzpicture}
  \gSetStdTypes{line}
}

\makeatother

\numberwithin{equation}{section}

\newcommand{\nn}{\nonumber}

\newcommand\beq{\begin{equation}}
\newcommand\eeq{\end{equation}}
\newcommand\beal{\begin{aligned}}
\newcommand\eeal{\end{aligned}}
\newcommand\bea{\begin{eqnarray}}
\newcommand\eea{\end{eqnarray}}

\newcommand\dd{{\mathrm d}}
\usepackage{xcolor}

\newcommand{\bp}{{\boldsymbol p}}
\newcommand{\bP}{{\boldsymbol P}}
\newcommand{\bL}{{\boldsymbol L}}
\newcommand{\bJ}{{\boldsymbol J}}
\newcommand{\bq}{{\boldsymbol q}}
\newcommand{\br}{{\boldsymbol r}}
\newcommand{\bR}{{\boldsymbol R}}

\newcommand{\bS}{{\boldsymbol S}}
\newcommand{\ba}{{\boldsymbol a}}

\newcommand\cE{\mathcal{E}}

\newcommand\cM{\mathcal{M}}
\newcommand\cS{\mathcal{S}}
\newcommand\cP{\mathcal{P}}

\makeatletter
\newcommand{\Biggg}{\bBigg@{3.5}}
\makeatother

\begin{document}
\preprint{DESY\,19-201\\\phantom{~} \hfill SLAC-PUB-17487\\ \phantom{~}\hfill UUITP-46/19}
\title{\center From Boundary Data to Bound States II:\\ [0.2cm] Scattering Angle to Dynamical Invariants (with Twist)}

\author[a,b]{\large Gregor K\"alin}
\author[c,d]{\large and Rafael A. Porto}
\affiliation[a]{SLAC National Accelerator Laboratory, Stanford University, Stanford, CA 94309, USA}
\affiliation[b]{Department of Physics and Astronomy, Uppsala University, Box 516, 751 20 Uppsala, Sweden}
\affiliation[c]{Deutsches Elektronen-Synchrotron DESY, Notkestrasse 85, 22607 Hamburg, Germany}
\affiliation[d]{The International Center for Theoretical Physics, Strada Costiera 11, Trieste 34151, Italy}
\emailAdd{gregor.kaelin@physics.uu.se,\,rafael.porto@desy.de}
\abstract{
  We recently introduced in [\href{https://arxiv.org/abs/1910.03008}{\texttt{1910.03008}}] a \emph{boundary-to-bound} dictionary between gravitational scattering data and observables for bound states of non-spinning bodies.
  In~this paper, we elaborate further on this \emph{holographic} map.
  We~start by deriving the following --- remarkably simple --- formula relating the periastron advance to the scattering~angle: $  \Delta \Phi(J,\cE) =\chi(J,\cE) + \chi (-J,\cE)$, via analytic continuation in angular momentum and binding~energy.
  Using explicit expressions from [\href{https://arxiv.org/abs/1910.03008}{\texttt{1910.03008}}], we confirm its validity to all orders in the Post-Minkowskian (PM)~expansion.
  Furthermore, we reconstruct the radial action for the~bound state directly from the knowledge of the scattering angle.
  The radial action enables us to write compact expressions for dynamical invariants in terms of the deflection angle to all PM orders, which can also be written as a function of the PM-expanded amplitude.
  As~an example, we reproduce our result in [\href{https://arxiv.org/abs/1910.03008}{\texttt{1910.03008}}] for the periastron advance, and compute the radial and azimuthal frequencies and redshift variable to two-loops.
  Agreement is found in the overlap between PM and Post-Newtonian (PN) schemes.
  Last but not least, we initiate the study of our dictionary including spin.
  We demonstrate that the same relation between deflection angle and periastron advance applies for aligned-spin contributions, with~$J$ the (canonical) \emph{total} angular momentum.
  Explicit checks are performed to display perfect agreement using state-of-the-art PN results in the literature. Using the map between test- and two-body dynamics, we also compute the periastron advance up to quadratic order in spin, to one-loop and to all orders in velocity.
  We conclude with a discussion on the generalized `impetus formula' for spinning bodies and black holes as `elementary particles'.
  Our findings here and in [\href{https://arxiv.org/abs/1910.03008}{\texttt{1910.03008}}] imply that the deflection angle already encodes vast amount of physical information for bound orbits, encouraging independent derivations using numerical and/or self-force methodologies.}

\maketitle
\newpage

\section{Introduction} \label{sec:sum}

Motivated by the new era of gravitational wave science dawning upon us \cite{GBM:2017lvd,LIGOScientific:2018mvr,Zackay:2019tzo,Venumadhav:2019lyq,Zackay:2019btq}, as well as the vast computational challenges~\cite{Porto:2016zng,Porto:2017lrn,review}, we have introduced in \cite{b2b1} (hereafter paper\,I) a \emph{boundary-to-bound} dictionary between gravitational scattering data and dynamical invariants for elliptic orbits.
We were able to bypass the need of rather lengthy and gauge dependent objects, e.g. the Hamiltonian, by directly mapping scattering information to gauge independent quantities for bound states, thus simplifying a key step required to construct accurate waveforms, while revealing a surprising connection between observables naturally defined at the \emph{boundary} and those in the \emph{bulk} of spacetime.
The construction in paper\,I was built upon a remarkable connection between the relative momentum of the two-body system and the scattering amplitude in the (conservative) classical limit, which we dubbed the \emph{impetus formula}.
For generic orbits, the latter allowed us to construct a radial action depending only on the --- analytically continued --- scattering amplitude, from which dynamical invariants such as the periastron advance ($\Delta\Phi$) can be obtained by differentiation.
Moreover, using Firsov's formula \cite{firsov,landau1,landau2} --- relating the scattering angle ($\chi$) to the distance of closest approach --- we have identified the orbital elements for elliptic orbits from hyperbolic motion via an additional analytic continuation in the impact parameter.
By imposing the vanishing of the eccentricity, we were able to simplify the derivation of the orbital frequency for circular orbits ($\Omega$) directly from scattering~data.
As~an example, we derived expressions for $\Omega$ and $\Delta\Phi$ directly from the knowledge of the scattering amplitude to two-loops, to all orders in velocity, reproducing known results to second order in the Post-Newtonian expansion (PN) while also providing a subset of  exact (`non-renormalized') contributions to all PN orders.
By resorting to a `no-recoil' approximation for the amplitude, together with the impetus formula, we also unveiled the reason behind the map between test-body and two-body dynamics to second Post-Minkowskian (2PM) order, originally discovered in~\cite{Vines:2018gqi}. \vskip 4pt

The purpose of this paper is to continue developing further the dictionary of paper\,I.
In~particular, we will elaborate on the (re-)construction of the radial action from boundary data, and the computation of all the independent gravitational observables for bound orbits.
In principle, the radial action was introduced, see e.g.~\cite{Damour:1988mr}, as an integral over the radial relative momentum of the two-body system.
Only after the impetus formula obtained in paper\,I is invoked, we were able to connect the latter to the scattering amplitude, thus opening the possibility to relate bound and unbound dynamics for generic orbits.
However, despite being gauge invariant, the classical limit of the scattering amplitude per se does not constitute an observable quantity.\footnote{
  Nevertheless, we have argued in paper\,I that the impetus formula invites itself to interpret the PM coefficients of the scattering amplitude (after a Fourier transformation into `coordinate space') as carrying physical information in the form of \emph{asymptotic~charges}.}
At the same time, in paper\,I we also derived an expression for the scattering angle as a function of the amplitude to all PM orders.
Therefore, it is natural to explore the possibility to recast our dictionary by re-expressing the coefficients of the amplitude in terms of those of the deflection angle, for instance for the computation of the periastron advance.
As we discuss here, a remarkable simplification arises, directly connecting the latter to the (even coefficients of the) former in the PM framework.
It turns out, once this relationship is found, the vestiges of the impetus formula disappear, hence begging for a more general explanation.
We provide it as the starting point of our paper, by demonstrating the following remarkably simply relationship 
\beq
\Delta\Phi(J,\cE) = \chi(J,\cE) + \chi(-J,\cE)\,, \qquad \cE<0\,, \label{eq1}
\eeq
between the periastron advance and the analytic continuation of the scattering angle, both in angular momentum and binding energy, in the conservative sector.
We provide not only the basis for the relation in \eqref{eq1} but also extensive evidence in concrete calculations.
In~particular, we demonstrate its validity in an exact case as well as to all orders in PM theory.\vskip 4pt

Armed with \eqref{eq1}, we can then reconstruct the radial action entirely in terms of the scattering angle, by integrating with respect to the angular momentum.
As we show, the integration \emph{constant} may be matched in the large angular momentum limit, which can be computed exactly.
Using the expressions derived in paper\,I, we then provide a compact expression in the PM framework that can be used to obtain all of the gravitational observables of the two-body system, directly from the (analytically continued) scattering angle.
As an example we derive the azimuthal and radial frequency as well as the redshift function to two-loop orders, in addition to (re-)deriving the expression for the periastron advance first obtained in paper\,I.\vskip 4pt

Due to several subtleties in the definition of the momentum and orbital elements, we have not attempted in this paper to study the extension of our formalism to spinning bodies with general orientations.
Yet, the fact that the assumptions leading to \eqref{eq1} are quite general, naturally led us to explore whether it applies once rotation is included, at least in some restricted situations.
We demonstrate that is indeed the case for black holes with aligned-spins.
The map consists on replacing $J$ in \eqref{eq1} by the total {\it canonical} angular momentum.
The proof of \eqref{eq1} for the case of aligned-spins relies solely on the existence of the quasi-isotropic gauge, in which the Hamiltonian depends on the canonical momentum via the combination $\bP^2 = P_r^2+L^2/r^2$, with $L$ the canonical orbital angular momentum, except in the odd-spin case where one has single factors of $\bL\cdot\ba \to L a$, with $\ba$ the spin parameter (with units of length) \cite{Vines:2018gqi}.
Even though at this stage we resort to the existence of the quasi-isotropic gauge, this is ultimately the one (implicitly) chosen by the Fourier transform of the amplitude in the center of mass frame.
Therefore, provided the matching discussed in \cite{ira1,Cheung:2018wkq,zvi1,zvi2} carries over to spin, as suggested in \cite{Vaidya:2014kza,Chung:2018kqs,Chung:2019duq,Arkani-Hamed:2019ymq,Guevara:2018wpp,Guevara:2019fsj,Guevara:2017csg,donalvines}, the existence of this gauge is guarantee to all PM orders.
As before, the exact form of the Hamiltonian is never needed, although it may be obtained and shown to agree with the existent literature, e.g. \cite{nrgrs,prl,nrgrso,nrgrs1,nrgrs2,Vines:2016qwa,Vines:2017hyw,levi}.
As an example, we explicitly show that \eqref{eq1}, applied to the results obtained in \cite{Vines:2018gqi} for the deflection angle, accurately predicts the value of the PN-expanded periastron advance to 3.5PN order computed in \cite{Sch12}.
Moreover, using the map between test- and two-body dynamics for spinning bodies to 2PM uncovered~in~\cite{Vines:2018gqi}, we compute the periastron advance up to quadratic order in the spin, to one-loop and to all orders in velocity.
We will return to the study of spin effects in forthcoming~work. 

\section{Radial Action} 

The classical problem of motion in gravity involving two non-rotating objects occurs in~a~plane, which we can choose to coincide with $\theta=\pi/2$. Following Hamilton-Jacobi
theory, and given the translational and rotational invariance in time and $\phi$, there exist an effective action describing the dynamics in the center of mass frame of the form
\beq
S= - \mu \cE t + J \phi + S_r(J,\cE),
\eeq
where the conservation of energy, 
\beq
E =M + \mu\cE = M(1+\nu \cE)\,,\label{eq:E}
\eeq
is manifest. Here $M=m_1+m_2$ is the total mass, $\mu = m_1m_2/M$ is the reduced mass, $\nu \equiv \frac{\mu}{M}$ is the symmetric mass ratio, and $J$ is the angular momentum. The \emph{radial action} takes the form \cite{Damour:1988mr}:
\beq
S_r = \frac{1}{2\pi} \oint p_r(J,\cE,r) \dd r\,,\label{eq:radial}
\eeq
where \beq p_r (J,\cE,r) \equiv \sqrt{\bp^2(\cE,r)-J^2/r^2}\,,\eeq is the radial momentum of the two-body system in the center of mass, written as a function of the energy and angular momentum, by solving for $p_r$ the equation $H(p_r,J,r)=E$.
Depending on the type of trajectory, unbound ($\cE>0$) or bound ($\cE<0$), the endpoints of the radial integral are the point of closest approach and infinity, as in a scattering process, or the motion occurs between the real positive zeros of $p_r$, as for elliptic orbits.
The scattering angle and periastron advance can be obtained by taking the derivative with respect to the angular momentum of the radial action. \vskip 4pt

Notice that, at this stage, we have not made any additional assumption about the motion, which is assumed to be conservative, other than the fact that the bodies are non-rotating.
As we shall see, continuing the development of the dictionary put forward in paper\,I, the scattering angle can be directly connected to the periastron advance for bound states, via analytic continuation in energy and angular momentum.
This can be achieved once the orbital elements are identified, as we did in paper\,I.\vskip 4pt In what follows we will denote as $\tilde r_\pm$ the roots associated to the radial variable for the scattering problem; and omit the tilde, e.g. simply $r_\pm$, for the case of bound orbits \cite{b2b1}. 
 
\subsection{Deflection Angle}
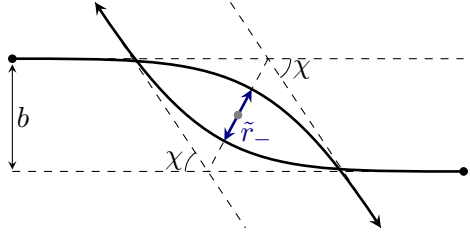
\begin{figure}
  \centering
  \begin{tikzpicture}
      [scale=0.5]
    \draw[dashed] (-6,1.5) -- (6,1.5);
    \draw[dashed] (-6,-1.5) -- (6,-1.5);
    \draw[dashed] (-3.8,3) -- (0.2,-3);
	\draw[dashed] (-0.2,3) -- (3.8,-3);
    \draw[dashed] (0.8,1.5) -- (-0.8,-1.5);
    \draw[->,line width=1] (-6,1.5) .. controls (-0.6,1.5) and (0.8,1.5) .. (3.8,-3);
    \draw[->,line width=1] (6,-1.5) .. controls (0.6,-1.5) and (-0.8,-1.5) .. (-3.8,3);
    \draw[<->,line width=1,black!50!blue] (-0.363,-0.68) -- node[very near start, right=0.1] {$\tilde{r}_-$} (0.363,0.68);
    \draw[<->] (-6,1.35) -- node[right] {$b$} (-6,-1.45);
    \draw (1.4,1.5) arc (0:-57:0.6);
    \node at (1.7,1.2) {$\chi$};
    \draw (-1.4,-1.5) arc (180:123:0.6);
    \node at (-1.7,-1.2) {$\chi$};
    \filldraw (-6,1.5) circle (0.1);
    \filldraw (6,-1.5) circle (0.1);
    \filldraw[gray] (0,0) circle (0.1);
  \end{tikzpicture}
  \caption{The geometry of the scattering problem. The motion of the bodies traces two hyperbolas, which are separated by $\tilde{r}_-$ at the point of closest approach. See paper\,I for details.}
  \label{fig:scat}
\end{figure}

The computation of the deflection angle is standard in the literature, see e.g. \cite{landau1,landau2}.
It can also be derived directly from the radial action, with a contour `around infinity',
\beq
S_r (J,\cE) = \frac{2}{2\pi} \int_{r_{\rm min}(J,\cE)}^\infty p_r (J,\cE,r) \dd r\quad \quad (\rm unbound)
\eeq
where $\cE>0$. (The factor of two is due to the `return trip'.)
The function $r_{\rm min}(J,\cE)>0$ is the point of closest approach in hyperbolic motion.
The impact parameter is related to the angular momentum via $J=p_\infty b$, with $p_\infty$ the center of mass momentum at infinity.
In the notation of paper\,I, we have
\beq
p_\infty^2 = \mu^2 \frac{\gamma^2-1}{\Gamma^2},
\eeq
with $\gamma = p_1\cdot p_2/(m_1m_2)$ in the center of mass and $\Gamma=E/M$.
We will also identify $r_{\rm min} = \tilde r_-$, the one positive  (real) root obeying \beq p_r(J,\cE,\tilde r_\pm(J,\cE))=0,\label{eqpr}\eeq with the other root, ${\tilde r}_+$, being negative (see Fig.~\ref{fig:scat}) \cite{b2b1}. Then, we have
\beq
\frac{\chi+\pi}{2\pi}   = -{\partial  S_r(J,\cE) \over \partial J} =  \frac{1}{\pi} \int_{{\tilde r}_-(J,\cE)}^\infty \,\, \frac{J}{r^2\sqrt{\bp^2(\cE,r)-J^2/r^2}}\dd r\,,
\eeq
for the scattering angle.

\subsection{Periastron Advance}
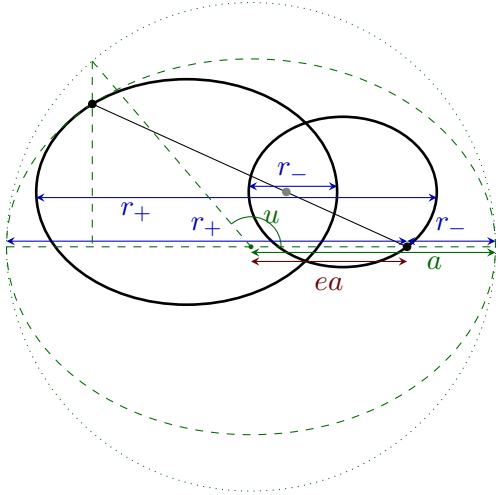
\begin{figure}
  \centering
  \begin{tikzpicture}
    [scale=0.5]
    \draw[line width=1] (0,0) ellipse (4 and 3);
    \draw[line width=1] (4.15,0) ellipse (2.5 and 2);
    \draw[] (-2.51,2.34) -- (5.86,-1.46);
    
    \draw[<->,black!30!blue] (-4,-0.15) -- node[below,near start] {$r_+$} (6.65,-0.15);
    \draw[<->,black!30!blue] (1.65,0.15) -- node[above] {$r_-$} (4,0.15);
    \draw[dashed,black!60!green] (1.71,-1.46) ellipse (6.5 and 5);
    \draw[dotted,black!60!green] (1.71,-1.46) circle (6.5);
    \draw[dashed,black!60!green] (-4.79,-1.46) -- (8.21,-1.46);
    \draw[<->,black!30!blue] (-4.79,-1.31) -- node[above] {$r_+$} (5.86,-1.31);
    \draw[<->,black!30!blue] (5.86,-1.31) -- node[above] {$r_-$} (8.21,-1.31);
    \draw[dashed,black!60!green] (-2.51,3.48) -- (-2.51,-1.46);
    \draw[dashed,black!60!green] (-2.51,3.48) -- (1.71,-1.46);

    \draw[<->,black!60!green] (1.71,-1.61) -- node[below,near end] {$a$} (8.21,-1.61);
    \draw[<->,black!60!red] (1.71,-1.85) -- node[below=0.15] {$e a$} (5.86,-1.85);

    \draw[black!60!green] (2.51,-1.46) arc (0:131:0.8);
    \node[black!60!green] at (2.25,-0.65) {$u$};

    \filldraw (-2.51,2.34) circle (0.1);
    \filldraw (5.86,-1.46) circle (0.1);
    \filldraw[gray] (2.65,0) circle (0.1);
    \filldraw[black!60!green] (1.71,-1.46) circle (0.05);
  \end{tikzpicture}
   \caption{Bound elliptic motion in the center of mass frame.  The black ellipses mark the individual paths. The heavier body lies on the focus of the green dashed ellipse. The latter describes the worldline of the lighter body in the companion's frame. The dotted circle of radius $a$ defines the eccentric anomaly, $u$, with $e$ the eccentricity. See paper\,I for details.}
  \label{fig:orbit}
  \end{figure}

Similar considerations apply to the bound case, except that now we have two real and positive roots in \eqref{eqpr}, denoted as $r_\pm$ in paper\,I, which determine the turning points of the orbit, see Fig.~\ref{fig:orbit}.
The radial action takes the form
\beq
S_r (J,\cE) = \frac{2}{2\pi}\int_{r_-(J,\cE)}^{r_+(J,\cE)} p_r (J,\cE,r) \dd r \, \quad \quad (\rm bound) \label{eq:sphi}
\eeq
with $\cE<0$, and the same factor of $2$ to complete the orbit. The periastron advance follows via differentiation
\beq
\frac{\Delta \Phi+2\pi}{2\pi} = - {\partial  S_r(J,\cE) \over \partial J} =  \frac{1}{\pi} \int_{r_-(J,\cE)}^{r_+(J,\cE)} \frac{J}{r^2\sqrt{\bp^2(\cE,r)-J^2/r^2}}\dd r\,.\label{partialJ}
\eeq
The reader will immediately notice that, written in this form, the similarity to the scattering angle is conspicuously displayed.

\subsection{Endpoints: Hyperbola vs Ellipse}\label{sec:orb}
As it was demonstrated in paper\,I, the orbital elements for  bound and unbound orbits can be related via analytic continuation.
For the scattering process, the point of closest approach~$r_{\rm min}$ corresponds to the positive of the two (real) roots of \eqref{eqpr}, while the other root is negative.
The two (real) roots for  bound orbits, $0 < r_- < r_+$, can then be obtained from to the hyperbolic case as follows. First the smaller root is obtained via analytic continuation in the energy and impact parameter \cite{b2b1}
\beq
 r_-(b,\cE)= \tilde r_-(ib,\cE)\, \qquad  b>0,\,\cE<0\,,\label{eqrpm2}
\eeq
where the (real) impact parameter is related to the angular momentum via
\beq
b= J/|p_\infty| >0\,. \label{eq:bn}
\eeq
Notice we have taken the absolute value of $p_\infty$, which is purely imaginary for negative energies, namely $p_\infty \to - i p_\infty$, such that $p_\infty^2 <0$ for bound orbits. 
In terms of the angular momentum, we have $J = p_\infty b = (-i p_\infty)( i b) > 0$, therefore it remains the same under the above analytic continuation. This implies
 \beq
 r_-(J,\cE) = \tilde r_- (J ,\cE)\, \qquad  J>0,\,\cE<0\,.
 \eeq
In other words, as expected from the condition in \eqref{eqpr}, one of the roots is simply related by analytic continuation in the energy at fixed (positive) angular momentum.\vskip 4pt
For the other root, which for the scattering problem is negative, we showed that it can be connected to $\tilde r_-$ via an additional analytic continuation in the impact parameter \cite{b2b1}, 
\beq
 r_+(b,\cE)= \tilde r_-(-ib,\cE)\, \qquad b>0,\,\cE<0\,,
\eeq
 or equivalently, using \eqref{eqrpm2}, 
\beq
r_+ (b,\cE)=  r_-(-b,\cE)\,, \qquad b>0\,,
\eeq
both evaluated at fixed (negative) binding energy. In terms of the angular momentum, the  analytic continuation to negative impact parameter and binding energy implies $J \to - J$, which yields
\beq
 r_+ (J,\cE) = \tilde r_-(-J,\cE) \qquad J>0,\, \cE<0\,,
\eeq
or equivalently, 
\beq
 r_+ (J,\cE)  = r_-(-J,\cE) \qquad J>0\,.
 \eeq
The above relationships will play a central role in connecting the scattering angle and periastron advance, as we show next. 
\section{From Scattering Angle to Periastron Advance $\dots$}

\subsection{Analytic Continuation}
The idea is to relate the radial action for the bound and unbound case, via analytic continuation.
In order to remove the upper limit at infinity, let us consider the following combination,
\beq
\left(\frac{\chi(J,\cE)}{2\pi} + \frac{1}{2}\right)+ \left(\frac{\chi(-J,\cE)}{2\pi} + \frac{1}{2}\right) = \frac{1}{\pi} \int_{\tilde r_-(J,\cE)}^{\tilde r_-(-J,\cE)} \frac{J}{r^2\sqrt{\bp^2(\cE,r)-J^2/r^2}}\dd r\,,\label{chi212}
\eeq
where we used that, without spins,  
\beq 
p_r(J,\cE,r) = p_r(-J,\cE,r)\,.
\eeq
(We will return to this condition in \S\ref{sec:spin} when we discuss spin effects.) Hence, from the analytical continuation to $\cE<0$, which connects the endpoints from hyperbolic to elliptic motion (see \S\ref{sec:orb}), we find
\beq
\beal
1+\frac{1}{2\pi} \left(\chi(J,\cE) + \chi(-J,\cE)\right) &=  \frac{1}{\pi} \int_{ r_-(J,\cE)}^{ r_+(J,\cE)} \frac{J}{r^2\sqrt{\bp^2(\cE,r)-J^2/r^2}}\dd r
\\ &= 1+ \frac{1}{2\pi} \Delta \Phi(J,\cE)\,,
\eeal
\eeq
such that
\beq
 \Delta \Phi(J,\cE) = \chi(J,\cE) + \chi(-J,\cE)\,,\qquad \cE<0\,,\label{eq:phichi}
\eeq
analytically continued both in angular momentum and binding energy.
As a side-product, the above relationship implies $\Delta \Phi(J,\cE) = \Delta \Phi(-J,\cE)$, which is indeed a symmetry in the conservative sector.\vskip  4pt

Notice that, while $\chi$ in principle is an incomplete hyper-elliptic integral, its symmetric part (in $J \to -J$) becomes a complete elliptic integral, which gives us the periastron advance after analytic continuation in the energy.
Since we have not assumed a perturbative expansion, the above relationship applies also in the non-perturbative (conservative) regime, including radiation-reaction effects (in the regime where the scattering angle is a smooth function of energy and angular momentum).

\subsection{Post-Minkowskian Expansion}
The non-perturbative relationship between scattering angle and periastron advance can also be studied in the PM framework.
Introducing the PM expansions in $j=J/(G M \mu)$ (notice only even terms in $j$ contribute for non-rotating bodies) 
\beq
{\Delta \Phi}(j,\cE)= \sum_{n=1} \Delta \Phi_j^{(2n)}(\cE)/j^{2n}\,, \label{eq:pmjp}
\eeq
and
\beq
\frac{\chi}{2}(j,\cE) = \sum_{n=1} \chi^{(n)}_j(\cE)/j^n\,,\label{eq:pmangle}
\eeq 
the map in \eqref{eq:phichi} yields, for non-spinning bodies,
\beq
  \Delta \Phi_j^{(2n)}(\cE)= 4\, \chi^{(2n)}_j(\cE)\,, \label{eq:pmpc} 
\eeq
after analytic continuation in the energy.\vskip 4pt

In paper\,I we provided expression for both the LHS and RHS of \eqref{eq:pmpc} in terms of the scattering amplitude, via the radial action together with the impetus formula \cite{b2b1}
\beq
\bp^2 = p_\infty^2 + \widetilde\cM(r,\bp)\,,  \label{impetus0} 
\eeq
(ignoring thus far radiation-reaction effects) expanded perturbatively as
\beq
= p_\infty^2\left( 1+ \sum_{i=1} f_i(\cE) \frac{(GM)^i}{r^i} \right) = p_\infty^2+ \sum_{n=1} {\widetilde\cM}_n(\cE) \frac{G^n}{r^n}  \label{impetus}\,, 
\eeq
where\footnote{${\cal M}(\bq,\bp) = \sum_n G^n {\cal M}_n(\bp,\bq)$ is the (IR-finite part of the) relativistic $2\to2$ amplitude in the (conservative) classical~limit. See paper\,I for more details.}
\beq
{\widetilde\cM}_n(\cE) = \frac{r^n}{2E} \int \frac{\dd^3\bq}{(2\pi)^3} {\cal M}_n(\bp,\bq) e^{i\bq\cdot\br} \,.
\eeq
For instance, using \eqref{partialJ} to 4PM order we found \cite{b2b1}
\beq \label{4pmperi}
\Delta\Phi (j,\cE) = \pi \frac{\widetilde\cM_2}{\mu^2M^2 j^2} +\frac{3\pi}{4}\frac{1}{M^4 \mu^4 j^4}\big(\widetilde\cM_2^2+2\widetilde\cM_1\widetilde\cM_3+2p_\infty^2\widetilde\cM_4\big)+\cdots \,. \quad 
\eeq    
On the other hand, we also derived the PM coefficients of the scattering angle as a function of the scattering amplitude,
\beq \chi_j^{(n)} = \hat p_\infty^n \chi_b^{(n)} = \hat p_\infty^{n} \frac{\sqrt{\pi}}{2} \Gamma\left(\frac{n+1}{2}\right)\sum_{\sigma\in\mathcal{P}(n)}\frac{1}{\Gamma\left(1+ \frac{n}{2}-\Sigma^{\ell}\right)}\prod_{\ell} \frac{f_{\sigma_{\ell}}^{\sigma^{\ell}}}{\sigma^{\ell}!}\,,  
\label{eq:phi2}
\eeq  where $\hat p_\infty= p_\infty/\mu$.
The above expression is written in terms of integer-partitions of $n=\sigma_\ell \sigma^\ell$ (summation), with $\Sigma_{\ell}\equiv \sum_{\ell} \sigma^{\ell}$, see paper\,I for more details.
Through \eqref{eq:phi2}, the relationship in \eqref{eq:pmpc} yields\footnote{For the case $2n=4$ we have four partitions in \eqref{eq:phi2}: $2n=4\cdot 1 = 2\cdot 2 = 3\cdot 1 + 1\cdot 1 = 1 \cdot 4$\,, such that
\bea
\Delta \Phi_j^{(4)}&=& 4\, \chi^{(4)}_j=2 \hat p_\infty^4 \sqrt{\pi}\,\Gamma\left(\frac{5}{2}\right)\left(\frac{1}{\Gamma(2)}\frac{f_4^1}{1!}+\frac{1}{\Gamma(1)}\frac{f_2^2}{2!}+\frac{1}{\Gamma(1)}\frac{f^1_1f^1_3}{1!1!}+\frac{1}{\Gamma(-1)}\frac{f_1^4}{4!}\right)\nn \\
&=&\hat p_\infty^4 \frac{3\pi}{2}\left(f_4+\frac{1}{2}f_2^2+f_1f_3\right)\nn\,.
\eea
Notice that, due to ${1 \over \Gamma(-1)} \to 0$, the $f_1^4$ term is absent.}
\beq
 \begin{aligned}
 \Delta \Phi_j^{(2)}  &=  4 \chi^{(2)}_j = \pi \hat p_\infty^2 \, f_2 = \pi \frac{ {\widetilde\cM}_2}{M^2\mu^2}\,,    \\
\Delta \Phi_j^{(4)} &= 4 \chi^{(4)}_j  =\,  \frac{3  \pi  \hat p_\infty^4}{4}   \left(  f_2^2 + 2 f_1f_3 +  2 f_4\right) = \frac{3\pi}{ 4M^4\mu^4} \left( {\widetilde\cM}_2^2 + 2{\widetilde\cM}_1{\widetilde\cM}_3 + 2 p_\infty^2 {\widetilde\cM}_4 \right) \,,
\end{aligned}
\eeq
to 4PM order, which are in perfect agreement with our previous result in \eqref{4pmperi}.
Note that the connection to the scattering angle explains the factor of $p_\infty^2$ in front of $\widetilde{M}_4$.\vskip 4pt

We emphasize that the relationship in \eqref{eq:pmpc} does not rely on the impetus formula, which is only used to relate $f_n$'s appearing on the left- and right-hand side of \eqref{eq:pmpc}, to the $\widetilde\cM_n$ expansion coefficients of the scattering amplitude.
As we demonstrate in Appendix~\ref{appA}, see also \S\ref{sec:rec}, the agreement between the periastron advance and scattering angle continues to all PM orders.
We have checked that the agreement continues also in the PN theory, see \S\ref{sec:spin} where we incorporate spin effects.  

\subsection{The Exact $f_2$-Theory}\label{f2t}

The relationship in \eqref{eq:phichi}, however, is valid also in the non-perturbative regime.
An illuminating example is given by the `$f_2$-theory', in which we set $f_{n \neq 2} =0$.\footnote{As expected there is no precession at 1PM from \eqref{eq:phichi}.}
It is easy to show that in this case we only have even PM contributions to the scattering angle, 
\beq
\begin{aligned}
  \chi_j^{(2n)}[f_2] &= \hat p_\infty^{2n}\frac{\sqrt{\pi} \Gamma\left(n+\frac{1}{2}\right)}{2\Gamma(n+1)} f_2^n,\quad n=1,2,\cdots\\
  \chi_j^{(2n+1)} [f_2] &= 0\,,
\end{aligned}
\eeq
obtained from \eqref{eq:phi2}. It is straightforward to perform the resummation, see paper\,I, and we found 
\beq
\frac{\chi[f_2]+ \pi}{2}= \frac{\pi}{2\sqrt{1- \frac{\hat p_\infty^2 f_2}{j^2}}}\,. 
\eeq
On the other hand, the periastron advance can be also computed exactly for the $f_2$-theory, directly from the radial action at one-loop order \cite{b2b1}.
We obtained in paper\,I,
\beq
 \frac{\Delta\Phi [f_2]}{2\pi} =    \left(\frac{1}{\sqrt{1- \frac{{\widetilde\cM}_2}{M^2 \mu^2 j^2}}}-1\right) \,.  
\eeq
Hence, using  $\hat p_\infty^2 f_2 = {\widetilde\cM}_2/(M^2\mu^2)$, we have
\beq
\Delta\Phi  (J, \cE) =  \chi  (J,\cE) + \chi (-J,\cE) = 2 \, \chi (J,\cE) \,, \eeq
confirming the non-perturbative relation between scattering angle and periastron advance in~\eqref{eq:phichi}.

\section{ $\dots$ to Dynamical Invariants}

The direct connection between $\chi$ and $\Delta \Phi$ allows us to compute one of the key gravitational observables for bound states, via a simple analytic continuation.
In order to obtain all of the other observables (for non-spinning bodies), we need to reconstruct the radial action.
This can be easily done by integrating the relationship in \eqref{eq:phichi}, and matching the integration constant in the $J \to \infty$ limit, as we discuss next.

\subsection{Reconstructing the Radial Action}\label{sec:rec}

In paper\,I we obtained the periastron advance by first computing the radial action, and afterwards performing the partial derivative w.r.t. the angular momentum, see \eqref{partialJ}.
Armed with the apsidal precession directly from the scattering angle, we can now proceed in the opposite direction, and construct the radial action via integration of the relationship in \eqref{partialJ} together with \eqref{eq:phichi} and \eqref{eq:pmpc}.
We will do this in the PM framework, where we have
\beq
\label{sralpha}
\frac{{\cal S}_r}{GM\mu} = -  \left(j + \frac{2}{\pi} \sum_n  \chi^{(2n)}_j({\cE}) \int \frac{dj}{j^{2n}}\right) + \alpha({\cE})\,, 
\eeq
with $\alpha(\cE)$ an integration constant.
We can easily determine $\alpha(\cE)$ by matching the above expression to our derivation in paper\,I of the radial action,
\beq
\cS_r(J,\cE)=- \sum_{n=0}^\infty \sum_{\sigma\in\cP(n)}
  \frac{(-1)^{\Sigma^{\ell}}\Gamma\left(\Sigma^{\ell} - \frac{1}{2}\right)}{2\sqrt{\pi}}\cS_{\left\{n+2\Sigma^{\ell},\Sigma^{\ell}\right\}}\big(A,B,C\big) \prod_{\ell} \frac{D_{\sigma_{\ell}}^{\sigma^{\ell}}}{\sigma^{\ell}!}\,.\label{eq:irS}\ 
\eeq
The master integrals, $\cS_{\left\{n+2\Sigma^{\ell},\Sigma^{\ell}\right\}}$, can be found in Appendix~\ref{appA}. The $A,B,C,D_n$ functions are given by:
\beq
\begin{aligned}
\label{abcd}
-A &= -p_\infty^2\,, \\
2B &= p_\infty^2 f_1 GM\,, \\
-C &= J^2\left(1- p_\infty^2 f_2(GM)^2/J^2\right)\,,\\
D_n &=  p_\infty^2 f_{n+2} (GM)^{n+2}\,. \\ 
\end{aligned}
\eeq

In general, the radial action takes the form \cite{Damour:1988mr,b2b1} 
\beq
\begin{aligned}
  \mathcal{S}_r  =&\left(\frac{B}{\sqrt{-A}} - \sqrt{-C}\right) +\cdots\,, 
  \label{eq:srd}
\end{aligned}
\eeq
with the ellipses including polynomials in $D_n$ whose coefficients are combinations of functions of the type (see Appendix~\ref{appA})
\beq
\frac{B^k}{C^{{1 \over 2}(m+k-1)}} \label{eqapp}\,,
\eeq 
with $(m,k)$ positive integers. Hence, taking the limit $J \to \infty$ in both expressions for the radial action leaves only the leading order term behind, yielding
\beq
\lim_{J\to \infty}  \mathcal{S}_r  = \frac{B}{\sqrt{-A}} - \sqrt{-C}  =  \frac{p_\infty^2 f_1 GM}{2\sqrt{-p_\infty^2}} - J  = -J + GM\mu\, \alpha(\cE) \,.
\eeq
The integration constant is then simply given by:
\beq
\alpha(\cE) =  
   \frac{\hat p_\infty^2}{ \sqrt{-\hat p_\infty^2}} \frac{f_1}{2} =  {\rm sg} (\hat p_\infty)\chi_j^{(1)}\,,
\eeq
where we introduced
\beq
{\rm sg} (\hat p_\infty) \equiv \frac{\hat p_\infty}{\sqrt{-\hat p_\infty^2}}= \frac{\hat p_\infty}{|\hat p_\infty|}\,,
\eeq
and used the 1PM result
\beq
\chi_j^{(1)} = \hat p_\infty \frac{f_1}{2}\,.
\eeq
From here we obtain the remarkably simple form of the radial action for the bound state,\beq
i_r(j,\cE) \equiv \frac{{\cal S}_r}{GM\mu} =  {\rm sg}(\hat p_\infty
)\chi^{(1)}_j(\cE) - j \left(1 + \frac{2}{\pi} \sum_{n=1}  \frac{\chi^{(2n)}_j({\cE})}{(1-2n)j^{2n}}\right)  \,,\label{eq:ir}
\eeq
directly via analytic continuation to $\cE <0$ in the PM coefficients of the scattering angle.\vskip 4pt

We can also write a compact formula in terms of the scattering amplitude to a given $n$PM order.
Using the expression for the scattering angle in terms of the $f_n$'s derived in paper\,I (see \eqref{eq:phi2}), we find\footnote{We have absorbed the $-j$ into the $n=0$ contribution in the sum.
  This follows after defining $\sum_{\sigma\in\cP(0)} \#=~1$, and noticing $\frac{j}{2\sqrt{\pi}} \Gamma\left(-\frac{1}{2}\right)=-j$.}
\beq
\begin{aligned}
\label{eq:irg}
  i_r (j,\cE) &= \frac{\hat p_\infty^2}{ \sqrt{-\hat p_\infty^2}} \frac{f_1}{2}
  + \frac{j}{2\sqrt{\pi}} \sum_{n=0}^\infty  \left(\frac{\hat p_\infty}{j}\right)^{2n}
  \Gamma\left(n-\frac{1}{2}\right)\sum_{\sigma\in\mathcal{P}(2n)}\frac{1}{\Gamma\left(1+ n-\Sigma^{\ell}\right)}\prod_{\ell} \frac{f_{\sigma_{\ell}}^{\sigma^{\ell}}}{\sigma^{\ell}!}\\
  &=  \frac{1}{2 \sqrt{-p_\infty^2}}\frac{{\widetilde \cM}_1}{M \mu} + \frac{j}{2\sqrt{\pi}} \sum_{n=0}^\infty \Bigg(  \frac{\Gamma\left(n-\frac{1}{2}\right)}{(\mu Mj)^{2n}}\,
  \sum_{\sigma\in\mathcal{P}(2n)}\frac{p_\infty^{2(n-\Sigma^{\ell})}}{\Gamma\left(1+n-\Sigma^{\ell}\right)}\prod_{\ell} \frac{{\widetilde \cM}_{\sigma_{\ell}}^{\sigma^{\ell}}} {\sigma^{\ell}!}\Bigg)\,,
\end{aligned}
\eeq
where in the second line we used the impetus formula \eqref{impetus} to relate the $f_n$'s to the ${\widetilde\cM}_n$'s.
For instance, to 4PM order we obtain 
\beq
\begin{aligned}
i_r (j,\cE) 
 &= -j + \frac{\hat p_\infty^2}{ \sqrt{-\hat p_\infty^2}} \frac{f_1}{2} +  \frac{\hat p_\infty^2}{2j} f_2 +  \frac{\hat p_\infty^4}{8j^3}\Big( f_2^2 +2 f_1f_3+2f_4 \Big) +\cdots \\
&=  -j +  \frac{1}{2 \sqrt{ -\hat p^2_\infty}}\frac{{\widetilde \cM}_1}{M \mu^2} +  \frac{1}{2j} \frac{{\widetilde \cM}_2}{M^2\mu^2} +  \frac{1}{8j^3}\frac{\left( {\widetilde \cM}_2^2 +2 {\widetilde \cM}_1{\widetilde \cM}_3+2 p_\infty^2 {\widetilde \cM}_4 \right)}{M^4\mu^4}+\cdots\,.\end{aligned} \label{irm4}
\eeq

\subsection{One-loop Resummation}\label{sec:one}

As for the exact $f_2$ theory in \S\ref{f2t}, we can also perform a partial PM resummation of the radial action in \eqref{eq:ir}.
This is relatively straightforward in terms of the scattering amplitude.
As we demonstrate in Appendix~\ref{appA}, the PM expansion of the radial action in terms of master integrals derived in paper\,I is equivalent to the expansion in \eqref{eq:ir}, or \eqref{eq:irg}, reconstructed from the scattering angle.
At the same time, as it was observed also in paper\,I, the expression in \eqref{eq:irS} naturally resums the one-loop contribution.
This can be easily seen already with the first term.
To one-loop order only the leading term $\cS_{\{q=0,m=0\}}$ contributes, and we find the result quoted in \eqref{eq:srd}, 
\beq
i_r = \frac{\widetilde\cM_1}{2M\mu^2\sqrt{-\hat p_\infty^2}}-j\sqrt{1-\frac{{\widetilde\cM}_2}{M^2\mu^2 j^2}} + \cdots\,,
\eeq
which accounts for many of the one-loop contributions at $n$PM in the series expansion of \eqref{eq:irg}.
As we discussed in paper\,I, this expression already includes non-perturbative information, both in the PM and PN expansions.
We may now go one step further, and resum all the contributions from the one-loop term in the radial action.\vskip 4pt

At higher orders, the radial action involves polynomials in $D_n$, whose coefficients depend on the combinations shown in \eqref{eqapp}. In terms of the amplitude, these take the form
\beq
\frac{\widetilde\cM_1^k}{\left((J/G)^2-\widetilde\cM_2\right)^{{1 \over 2}(m+k-1)}}\,.
\eeq
Notice this naturally resums all the $1/j^n$ contributions from the one-loop term.
For example, keeping the leading terms in the `$D_{1,2}$-theory', but resuming all the one-loop contributions, we arrive at
\beq
i_r = \frac{\widetilde\cM_1}{2M\mu^2\sqrt{-\hat p_\infty^2}}-j\sqrt{1-\frac{{\widetilde\cM}_2}{M^2\mu^2 j^2}} \left(1- \frac{1}{\left(1 -\frac{\widetilde\cM_2}{M^2\mu^2 j^2}\right)^{2}}\left(\frac{ \widetilde\cM_1 \widetilde\cM_3+p_\infty^2{\widetilde \cM}_4}{4 M^4\mu^4j^4} \right)\right)+\cdots \,, \label{iresum}
\eeq
which recovers the expression in \eqref{irm4} after expanding in $1/j$.
In what follows we show how all the dynamical invariants of the bound state can be obtained directly from variations of the radial action, which itself may be obtained directly in terms of the scattering angle as we have shown here, or the scattering amplitude as discussed in paper\,I.

\subsection{Gravitational Observables}
We have already shown how the periastron advance follows from the scattering angle~in~\eqref{eq:phichi}.  Yet, armed with \eqref{eq:ir} we can also obtain, after analytic continuation in the energy, all the other observables for the two-body problem via differentiation.

\subsubsection*{Periastron Advance}
By construction, we have
\beq
\frac{\Delta\Phi(j,\cE)}{2\pi} = -  {\partial  \over \partial j} n(j,\cE) = \frac{1}{\pi} \sum_{n=1}  \frac{2\chi_j^{(2n)}(\cE)}{j^{2n}} = \frac{1}{\pi} \frac{\chi(J,\cE) + \chi(-J,\cE)}{2}\,,
\eeq
where $n=i_r+j$ is the so-called Delaunay variable, see e.g.  \cite{9912}. 

\subsubsection*{Azimuthal and Radial Frequencies}

The periastron-to-periastron period is given by
\beq
\begin{aligned}
\frac{T_p}{2\pi} = GM {\partial  \over \partial \cE} i_r(j,\cE) &=  GM \left( \partial_{\cE}\,\Big({\rm sg} (\hat p_\infty) \chi_j^{(1)}(\cE)\Big) -  \frac{2}{\pi} \sum_{n=1}  \frac{\partial_\cE\, \chi^{(2n)}_j({\cE})}{(1-2n)j^{2n-1}}\right)\\
&=  GE  \left( \partial_\gamma \Big({\rm sg} (\hat p_\infty) \chi_j^{(1)}(\gamma)\Big) -  \frac{2}{\pi} \sum_{n=1}  \frac{\partial_\gamma \, \chi^{(2n)}_j({\gamma})}{(1-2n)j^{2n-1}}\right)\,,
\end{aligned}
\eeq
where in the last line we used ${\partial \gamma \over \partial \cE} = \Gamma = E/M$.
From here we can also compute the radial and periastron frequencies~\cite{9912}
\beq
\Omega_r (j,\cE)\equiv \frac{2\pi}{T_p}\,, \quad
\Omega_p (j,\cE) \equiv \frac{\Delta\Phi}{T_p}\,,
\eeq
as well as the azimuthal frequency \cite{9912}, 
\beq
\Omega_{\phi} \equiv \Omega_r +\Omega_p= \frac{2\pi}{T_p} \left(1+\frac{\Delta\Phi}{2\pi}\right)\,.
\eeq
Alternatively, it can also be read-off directly from the radial action,
\beq
GM\Omega_\phi =  - \frac{1}{\Gamma} {\partial i_r  \over \partial j}\left({\partial i_r  \over \partial \gamma}\right)^{-1} \,.
\eeq
In terms of the scattering angle we find 
\beq
x= x_{\rm 1PM} \frac{\left(1+\frac{2}{\pi} \sum_{n=1}  \frac{\chi_j^{(2n)}(\gamma)}{j^{2n}}\right)^{2/3}}{\left(1 - x^{3/2}_{\rm 1PM}\,  \frac{2\Gamma}{\pi } \sum_{n=1}  \frac{\partial_\gamma \, \chi^{(2n)}_j({\gamma})}{ (1-2n)j^{2n-1}}\right)^{2/3}} \,,
\eeq
where we introduced the standard PN parameter $x \equiv (GM\Omega_\phi)^{2/3}$, and the 1PM contribution is given by\footnote{
  Notice that in the PN expansion we have, with $\epsilon = - 2\cE$, \beq x_{\rm 1PM} =\epsilon \left(1 + \frac{1}{12}\left(-15+\nu\right)\epsilon + \frac{1}{72}(180+15\nu+4\nu^2)\epsilon^2+\cdots\right) \,.
\eeq
As we already showed in paper\,I, the compact expression in \eqref{eq:x1pm} incorporates the exact ${\cal O}(\nu^n \epsilon^n)$ contributions to all PN orders.
(This is not surprising, since $f_1$ controls the 1PM theory.)
We can also see from here why the higher powers of $\nu$ are \emph{protected}.
This is simply due to the scaling $1/j^2\sim \epsilon$, as well as the smoothness of the $\nu \to 0$ limit.}
\beq
x_{\rm 1PM}  \equiv \frac{1}{ \,\Big(\Gamma \partial_\gamma \Big({\rm sg} (\hat p_\infty)\chi_j^{(1)}(\gamma)\Big)\Big)^{2/3}} =  \frac{(1-\gamma^2)}{\left(\Gamma(3\gamma-2\gamma^3)\right)^{2/3}}\,.  \label{eq:x1pm}
\eeq

\subsubsection*{Orbital Frequency for Circular Orbits}

For the case of circular orbits the azimuthal frequency turns into the orbital frequency, i.e. $\Omega_{\phi} \to \Omega_{\rm circ}$.
However, the derivation as a function of the binding energy only, still requires knowledge of the function $j(\cE)$.
This can be obtained by setting $i_r=0$ and solving for the angular momentum in \eqref{eq:ir}.
Notice, however, this is rather cumbersome in general.
Alternatively, as we explained in paper\,I\,, the function $j(\cE)$ can be derived directly through the determination of the orbital elements, and the condition of vanishing eccentricity.
Furthermore, once $j(\cE)$ is known, the orbital frequency also follows from the first law of binary dynamics \cite{alt2}, obtaining \cite{b2b1}
\beq
\Omega_{{\rm circ}} = \left({dj(\cE) \over d \cE}\right)^{-1} = \frac{1}{\Gamma} \left({dj(\cE) \over d \gamma}\right)^{-1}\,.
\eeq
Therefore, while equivalent to setting the radial action to zero, the analysis in paper\,I enforcing the vanishing of the eccentricity simplifies the derivation of the orbital frequency for circular orbits.

\subsubsection*{Redshift}

The first law of black hole dynamics states \cite{alt2}, in our language,\beq
\delta {\cal S}_r(J,E,m_a) =  - \left(1+\frac{\Delta\Phi}{2\pi}\right)\delta J + \frac{1}{\Omega_r} \delta E - \sum_{a=1,2} \frac{\langle z_a\rangle}{\Omega_r} \delta m_a\,,
\eeq
 where $\langle z_a\rangle$ is the averaged redshift. For constant mass, $\delta m_a=0$, the derivative with respect the total and binding energy coincide (see \eqref{eq:E}). However, when we allow for variations of the masses, the first law becomes
\beq
\delta \cS_r(J,\cE,m_a) = - \left(1+\frac{\Delta\Phi}{2\pi}\right)\delta J 
+ \frac{\mu}{\Omega_r}\delta\cE - \sum_a \frac{1}{\Omega_r}\left(\langle z_a\rangle-{\partial E(\cE,m_a) \over \partial m_a}\right)\delta m_a\,.
\eeq
 The derivatives with respect to $J$ and $\cE$ gave us the periastron advance and radial frequency, while the remaining one yields
 \beq
 \beal
\langle z_a\rangle -\langle z_a^{(0)}\rangle &=  -\Omega_r \frac{\partial {\cal S}_r}{\partial m_a}  = -\mu \left(\frac{\partial \cS_r}{\partial \cE}\right)^{-1} \frac{\partial {\cal S}_r}{\partial m_a}\\
&= -\frac{1}{M} \left( \partial_\cE \Big({\rm sg} (\hat p_\infty) \chi_j^{(1)}(\cE)\Big) -  \frac{2}{\pi} \sum_{n=1}  \frac{\partial_\cE \, \chi^{(2n)}_j({\cE})}{(1-2n)j^{2n-1}}\right)^{-1}\times \\
&\quad  \times \frac{\partial}{\partial m_a} \left[  m_1m_2 \left(\frac{\hat p_\infty}{\sqrt{-\hat p_\infty^2}} \chi^{(1)}_j(\cE,m_a) - \frac{2}{\pi} \sum_{n=1}  \frac{\chi^{(2n)}_j({\cE,m_a})}{(1-2n)}\left(\frac{Gm_1m_2}{J}\right)^{2n-1}\right) \right]
\eeal
 \eeq
 for the shift in the average redshift, with
 \beq
\langle z_a^{(0)}\rangle \equiv  {\partial E(\cE,m_a) \over \partial m_a} = 1+\frac{\partial \mu}{\partial m_a} \cE \,.
 \eeq
 
\subsection{$\{\Delta \Phi, \Omega_r,\Omega_\phi, z_1\}$ to Two-Loops} \label{sec:2loop}

Using the general formulas in \eqref{eq:ir} and \eqref{eq:irg} we can now construct all dynamical invariants using the state-of-the-art knowledge of the scattering angle to two-loops.
The relevant scattering amplitude was computed in \cite{zvi1,zvi2}, leading to the PM coefficients
\beq
\begin{aligned}
{\widetilde \cM}_1 &= 2 M \mu^2\left( \frac{2\gamma^2-1}{\Gamma}\right) \\
{\widetilde \cM}_2 &= \frac{3M^2\mu^2}{2} \left(\frac{5\gamma^2-1}{\Gamma}\right) \\
{\widetilde \cM}_3&=  -\frac{M^3\mu^2}{6\Gamma} \Bigg( 3-54\gamma^2 - 48 \nu(3+12\gamma^2-4\gamma^4)\frac{\arcsin\sqrt{\frac{1-\gamma}{2}}}{\sqrt{1-\gamma^2}}\\ &\quad+ \nu \left(-6+206 \gamma + 108 \gamma^2+4\gamma^3-\frac{18\Gamma(1-2\gamma^2)(1-5\gamma^2)}{(1+\Gamma)(1+\gamma)} \right)\Biggg)\,. \label{cMt}
\end{aligned}
\eeq
We can also perform a partial PM resummation of one-loop effects, as shown in \eqref{iresum}.
However, as we shall see below, the power counting is subtle.
As we noticed already in paper\,I with the derivation of the scattering angle and periastron advance from the amplitude, the different loop orders are mixed in the $1/j$ expansion.
Therefore, we will only keep terms up to two-loops, which is a consistent truncation in the PN expansion due to the factor of $p_\infty^2 \sim \cE$ in front of the three-loop amplitude in \eqref{eq:irg}.
Nevertheless, the structure of the expansion for the radial action strongly encourages the need of the three-loop contribution, in order to complete the $1/j^4$ term to all orders in velocity. 

\subsection*{Radial Action}
The radial action follows directly form \eqref{eq:irg}, using \eqref{cMt} we have
\beq
\begin{aligned}
i^{(L=2)}_r (j,\cE) = &-j +\frac{2\gamma^2-1}{\sqrt{1-\gamma^2}} +  \frac{3}{4 j}  \frac{5\gamma^2-1}{\Gamma}+  \frac{9}{32j^3}\frac{(5\gamma^2-1)^2}{\Gamma^2}  -\frac{1}{12 j^3}  \frac{2\gamma^2-1}{\Gamma^2} \Bigg( 3-54\gamma^2 \\ &+ \nu \left(-6+206 \gamma + 108 \gamma^2+4\gamma^3-\frac{18\Gamma(1-2\gamma^2)(1-5\gamma^2)}{(1+\Gamma)(1+\gamma)} \right)\\
&- 48 \nu(3+12\gamma^2-4\gamma^4)\frac{\arcsin\sqrt{\frac{1-\gamma}{2}}}{\sqrt{1-\gamma^2}}\Biggg)  
\end{aligned}
\eeq
From here it is straightforward to derive the gravitational observables for bound states, as discussed above.
\subsubsection*{Periastron Advance}
\beq
\beal
\frac{\Delta \Phi_{(L=2)}}{2\pi} = -\frac{\partial n_{(L=2)}}{\partial j} &= \frac{3}{4j^2} \frac{5\gamma^2-1}{\Gamma}  +  \frac{27}{32j^4}\frac{(5\gamma^2-1)^2}{\Gamma^2}  -\frac{1}{4 j^4}  \frac{2\gamma^2-1}{\Gamma^2} \Bigg( 3-54\gamma^2  \\ &+ \nu \left(-6+206 \gamma + 108 \gamma^2+4\gamma^3-\frac{18\Gamma(1-2\gamma^2)(1-5\gamma^2)}{(1+\Gamma)(1+\gamma)} \right)\\
&- 48 \nu(3+12\gamma^2-4\gamma^4)\frac{\arcsin\sqrt{\frac{1-\gamma}{2}}}{\sqrt{1-\gamma^2}}\Biggg) \label{peri3pm}\,,   
\eeal
\eeq
with $n_{(L=2)} = i_r^{(L=2)}+j$, is the Delaunay variable to two-loops.
Needless to say, this recovers the original expression derived in paper\,I.
\subsubsection*{Radial and Azimuthal Frequencies}
Up to two loops, expanding in $\epsilon = -2\cE$, we find
\beq
\begin{aligned}
  \frac{GM \Omega^{(L=2)}_r}{\epsilon^{\frac{3}{2}}} &= \textcolor{blue}{1-
   \frac{(15-\nu)}{8}\epsilon+\frac{555+30\nu+11\nu^2}{128 } \epsilon^2} \\
  &+ \left(\textcolor{blue}{\frac{3(2\nu-5)}{2 j}}\textcolor{red}{-\frac{194-184\nu+23\nu^2}{4  j^3}}\right)\epsilon^{\frac{3}{2}}\\
  &+\left(\textcolor{blue}{\frac{15(17-9\nu+2\nu^2)}{8 j}}\textcolor{cyan}{+\frac{21620-28592\nu+8765\nu^2-865\nu^3}{80  j^3}}\right)\epsilon^{\frac{5}{2}}
  + \cdots  \,,
  \end{aligned}
  \eeq
  \beq
  \begin{aligned}
  \frac{G M \Omega^{(L=2)}_\phi}{\epsilon^{\frac{3}{2}}}
  &=\textcolor{blue}{1+\frac{3}{j^2}-\frac{15(2\nu-7)}{4j^4}}+\left(\textcolor{blue}{\frac{1}{8}(\nu-15)+\frac{15(\nu-5)}{8j^2}} \textcolor{red}{-\frac{3(1301-921\nu+102\nu^2)}{32j^4}}\right)\epsilon\\
  &+\left(\textcolor{blue}{\frac{3(2\nu-5)}{2j}}\textcolor{red}{+\frac{-284+220\nu-23\nu^2}{4j^3}}\textcolor{cyan}{+\frac{3(913-728\nu+106\nu^2)}{j^5}}\right)\epsilon^{\frac{3}{2}}
\\
  &+\left(\textcolor{blue}{\frac{1}{128}(555+30\nu+11\nu^2)+\frac{3(895-150\nu+51\nu^2)}{128j^2}}\right.\\
&- \left. \textcolor{cyan}{\frac{3(-270085+251236\nu-70545\nu^2+7470\nu^3)}{2560j^4}}\right)\epsilon^2\\
  &+\left(\textcolor{blue}{\frac{15(17-9\nu+2\nu^2)}{8j}}
  \textcolor{cyan}{+\frac{31520-34442\nu+10025\nu^2-865\nu^3}{80j^3}}\right)\epsilon^{\frac{5}{2}}\,.\label{eq:omegas}
  \end{aligned}
\eeq
The color coding indicates the terms that agree with known results to 3PN order (\textcolor{blue}{blue}) obtained in \cite{alt2}, disagree at 3PN (\textcolor{red}{red}), and are higher order in velocity (\textcolor{cyan}{cyan}).
The disagreement is expected given the 3PM level of accuracy for the amplitude.
Note however, as we explained in paper\,I, that certain terms at 3PN are also recovered from the 3PM scattering angle, e.g. at ${\cal O} (\epsilon^2/j^2)$.
This is due to the fact that these are controlled by the one-loop term, and are not `renormalized' by higher order PM contributions \cite{b2b1}.

\subsubsection*{Redshift}
Using the two-loop results, we obtain for the redshift function
\beq
\begin{multlined}
  \langle z^{(L=2)}_2 \rangle = \textcolor{blue}{1+
  \frac{1}{4}(2\nu-3\Delta-3)\epsilon
  +\left(-\frac{3(1+\Delta)}{j}+\frac{5((5\nu-14)(1+\Delta)-4\nu^2)}{4j^3}\right)\epsilon^{\frac{3}{2}}}\\
  \textcolor{blue}{+\frac{1}{16}(3(10-\nu)(1+\Delta)+4\nu^2)\epsilon^2}
  +\left(\textcolor{blue}{-\frac{3(11\nu-35)(\Delta+1)-8\nu^2)}{8j}}\right.\\
  \left.\textcolor{red}{+\frac{(3378-3021\nu)(\Delta+1)+2165\nu^2+393\Delta\nu^2-388\nu^3}{32j^3}}\right)\epsilon^{\frac{5}{2}}\\
  \left(\textcolor{blue}{\frac{1}{32}(-(3\nu^2+130)(\Delta+1)+4\nu^3)-\frac{9(1+\Delta)(2\nu-5)}{2j^2}}\right.\\
 \left. \textcolor{cyan}{+\frac{3((738-633\nu)(\Delta+1)+196\nu^2+96\Delta\nu^2-4\nu^3)}{8j^4}}\right)\epsilon^3\,,
\end{multlined}
\eeq
where $\Delta\equiv \sqrt{1-4\nu}$, assuming without loss of generality that $m_1 \geq m_2$.
This factor accounts for the mass difference (recall $\nu=1/4$ for equal mass).
The value for $\langle z_1\rangle$ is obtained by the replacement $\Delta\rightarrow -\Delta$.
We have used the same color coding as in~\eqref{eq:omegas}. Notice the redshift function matches up to ${\cal O}(\epsilon^3)$ the value in \cite{alt2}. That is expected, since the Newtonian 0PN result starts at ${\cal O}(\epsilon)$.

\section{Aligned-Spins}\label{sec:spin}

The inclusion of spin effects introduces several subtleties, most notably the precession of the angular momentum.
Subsequently the motion is not restricted to a plane.
One can assume, however, that the spin and angular momentum are aligned throughout the evolution of the binary system.
This condition also requires the individual spins to be aligned.
Under these circumstances the dynamics of the bodies remains in a plane.
As we shall see, and much as we did for non-rotating bodies, the contribution from aligned-spin terms entering in the scattering angle for hyperbolic motion can be directly mapped to contributions to the periastron advance for bound states.
\vskip 4pt

In this section we will denote as $L$ the \emph{canonical} orbital angular momentum, to distinguish it from $J$, which in this section we will reserve for the total angular momentum, including the spin.
We will denote the canonical linear momentum as $\bP$, as opposite to $\bp$, which we use for the {\it physical} momentum, see e.g.~\cite{Vines:2018gqi}. To avoid introducing too many new variables, we will keep $\br$ as the position coordinate associated with the canonical momentum, unless otherwise noted. For the spin parameters we will use the standard vector $\ba_i$ (with unit of length) such that $\bS_i = m_i \ba_i$ (in $c=1$ units).
Moreover, we will present results as a function of $\tilde a_\pm \equiv a_\pm/(GM)$, with $ a_\pm =  a_1 \pm  a_2$ projected onto the spin axis, which coincides with that of the angular momentum.
As before, we will quote PN results in terms of $\epsilon= -2\cE \sim v^2$.

\subsection{Scattering Angle to Periastron Advance}

Let us return to the contour integral defined in \eqref{chi212}, but now shift the sign of the \underline{total} angular momentum, namely $ \bL \to -\bL$ {\it and} $\ba_i \to -\ba_i$, such that 
\beq
\bJ = \bL + m_1\ba_1 + m_2 \ba_2 \to -\bL - m_1\ba_1 - m_2 \ba_2\,.
\eeq
Hence, the same routing of the radial action we performed for the non-spinning case yields in this case
\beq
\beal
1+\frac{\chi(\bL,\ba_i,\cE)+\chi(-\bL,-\ba_i,\cE)}{2\pi} &= \frac{1}{\pi} \int_{\tilde r_-(J,\cE)}^{r_\Lambda} \frac{\dd r}{2\sqrt{P_r^2(\bL,\ba_i,\cE,r)}}\frac{\partial P^2_r}{\partial L}(\bL,\ba_i,\cE,r)\\
& +\frac{1}{\pi} \int_{\tilde r_-(-J,\cE)}^{r_\Lambda} \frac{\dd r}{2\sqrt{P_r^2(-\bL,-\ba_i,\cE,r)}}\frac{\partial P^2_r}{\partial L} (-\bL,-\ba_i,\cE,r)\,,
\eeal
\eeq
written in terms of $\bP$, the canonical linear momentum.
Inhere the $r_\Lambda$ is an (infrared) cutoff that can be taken to infinity at the end of the process.
Hence, provided 
\beq
P_r^2(\bL,\ba_i,\cE,r) = P_r^2(-\bL,-\ba_i,\cE,r)\,, \label{eqprl}
\eeq 
and
\beq
\frac{\partial P^2_r}{\partial L}(\bL,\ba_i,\cE,r) = -\frac{\partial P^2_r}{\partial L}(-\bL,-\ba_i,\cE,r)\,, \label{eqprl2}
\eeq
we have
\beq
1+\frac{\chi(J,\cE)+\chi(-J,\cE)}{2\pi} = \frac{1}{\pi} \int_{\tilde r_-(J,\cE)}^{\tilde r_-(-J,\cE)} \frac{1}{2\sqrt{P_r^2(\bL,\ba_i,\cE,r)}}\frac{\partial P^2_r}{\partial L}(\bL,\ba_i,\cE,r)\dd r\,,\label{chiL}
\eeq
and we can safely take $r_\Lambda \to \infty$.\vskip 4pt
The conditions in \eqref{eqprl}-\eqref{eqprl2} may not be satisfied in general.
However, as it was demonstrated in \cite{Vines:2018gqi}, the existence of a quasi-isotropic gauge guarantees that the aligned-spin Hamiltonian, $H(\br, \bP,\bS_1,\bS_2)$, only depends on the momentum via the combination
\beq
\bP^2 = P_r^2 + \frac{L^2}{r^2}\,,
\eeq 
except for odd-spin terms, where one has single factors of $\bL\cdot\ba_\pm = L a_\pm$.
Moreover, since both spin-spin and spin-orbit contributions are invariant under $J \to -J$,\footnote{
  This is guaranteed by time-reversal invariance, which holds as long as we work in the conservative sector.
  Notice that this in principle may also include radiation-reaction terms, see e.g.~\cite{tail,natalia1,natalia2,apparent,lamb,review,nrgr4}.}
the conditions in \eqref{eqprl}-\eqref{eqprl2} are obeyed.\vskip 4pt 

At the same time, the analysis in paper\,I of the orbital elements can be easily extrapolated to the case of aligned-spins, provided we use the ${\it canonical}$ impact parameter $b_{\rm can}\equiv L/p_\infty$.
The existence of a quasi-isotropic gauge implies that, in the PM framework, 
\beq
P_r^2(\cE,L,a_1,a_2) =  p^2_\infty \Bigg( 1 + \sum_i f_i(\cE,a^2_-,a^2_-, L a_+, L a_-)\frac{(GM)^i}{r^i}- \frac{b_{\rm can}^2}{r^2}\Bigg)\,,\label{eq:Pr}
\eeq
where we used that $P_\infty  = p_\infty$.\footnote{
  Notice that the canonical momentum, $\bP$, in general differs from the physical momentum, $\bp$, other than at infinity, see \cite{Vines:2018gqi}.}
Hence, we can now follow the same steps as in paper\,I.
For the first root, it is straightforward to show
\beq
r_-(J,\cE) = \tilde r_-(J,\cE)\,, \qquad \cE<0\,,
\eeq
since this solves $P_r=0$, with $\cE<0$ the condition for bound states.
The tricky part is to find the other solution, once spin is included.
However, it is also easy to see that 
\beq
 r_+(J,\cE) = \tilde r_-(-J,\cE)\,,
\eeq
remains valid.
That is clearly the case for even-spin terms, since the $a_\pm^2$ contributions go for the ride.
For the odd- and aligned-spin corrections, we notice that $\bL\cdot\ba \to La$ remains invariant under $L \to -L$ and $a\to -a$, and therefore we can follow the steps in paper\,I, with $b_{\rm can}$ as an independent variable in \eqref{eq:Pr}, and $La$ serving as a spectator inside the $f_i$'s, much like spin-spin terms.
We will return to the study of the orbital elements with spin term elsewhere.\vskip 4pt

We are now in a position to show that the expression in \eqref{chiL} yields, in the conservative sector,
\beq
\frac{\Delta\Phi(J,\cE)}{2\pi}=\frac{\chi(J,\cE)+\chi(-J,\cE)}{2\pi}\,, \qquad \cE<0 \,,
\label{eq:chiphiS}
\eeq
with $J$ the total angular momentum, as advertised. In what follows we confirm its validity in the framework of the PN expansion.

\subsection{Post-Newtonian Expansion}

The scattering angle for aligned-spins was computed in \cite{Vines:2018gqi} as a function of the energy and impact parameter, using the conservative PN Hamiltonian up to 3.5PN order with spin-orbit and spin-spin couplings \cite{Jaranowski:1997ky,Blanchet:2000ub,review,levi}.
The results in \cite{Vines:2018gqi} are given as a function of the (covariant) impact parameter~$b$, and relative velocity defined through $\gamma = 1/\sqrt{1-v^2}$.
In order to re-write the scattering angle as a function of the canonical orbital angular momentum, $L$, one uses \cite{Vines:2018gqi} 
\beq
L = p_\infty b + M \frac{\Gamma-1}{2}\left(a_+ - \frac{\Delta}{\Gamma} a_-\right)\,,\label{eqLb}
\eeq
which introduces spin-dependent terms also in the spin-independent contributions.\footnote{
  Notice that, for aligned-spins, the two spin supplementarity conditions (covariant and canonical) \cite{review}, lead to the same spin components orthogonal to the plane.
  Therefore, only the orbital part is shifted by the change of variables, see \cite{Vines:2018gqi}.}
Putting it all together, we arrive at:
\beq
\begin{aligned}
  \frac{\chi(\ell,a,\epsilon)}{2\pi} &=
  \left[\frac{1}{\pi}(-\epsilon)^{-\frac{1}{2}}
    -\frac{(\nu-15)}{8\pi}(-\epsilon)^{\frac{1}{2}}
    +\frac{35+30\nu+3\nu^2}{128\pi}(-\epsilon)^{\frac{3}{2}}\right]\frac{1}{\ell}\\
  &\begin{multlined}
     +\bigg[\textcolor{blue}{3
     +\frac{3(2\nu-5)}{4}\epsilon
     +\frac{3(5-5\nu+4\nu^2)}{16}\epsilon^2}
     -\frac{7 \tilde{a}_++\Delta\tilde{a}_-}{2\pi}\epsilon^{-\frac{1}{2}}\\
     +\frac{5\Delta(\nu-3)\tilde{a}_-+(23\nu-25)\tilde{a}_+}{16\pi}(-\epsilon)^{\frac{3}{2}}
     \bigg]\frac{1}{2\ell^2}
   \end{multlined}\\
  &\begin{multlined}
     +\bigg[\textcolor{blue}{-\frac{7\tilde{a}_++\Delta\tilde{a}_-}{2}
     -\frac{(\nu-6)\Delta\tilde{a}_-+(7\nu-18)\tilde{a}_+}{2}\epsilon}\\
     \textcolor{blue}{-\frac{3\left((15-14\nu+2\nu^2)\Delta\tilde{a}_-+(25-38\nu+14\nu^2)\tilde{a}_+\right)}{16}\epsilon^2}
    \\ -\frac{2}{3\pi}(-\epsilon)^{-\frac{3}{2}}
     +\frac{33+\nu}{4\pi}(-\epsilon)^{-\frac{1}{2}}
     +\frac{3003-1090\nu-5\nu^2+128\tilde{a}_+^2}{64\pi}(-\epsilon)^{\frac{1}{2}}
     \bigg]\frac{1}{2\ell^3}
   \end{multlined}\\
  &\begin{multlined}
    + \bigg[\textcolor{blue}{\frac{3(35+2\tilde{a}_+^2-10\nu)}{4}
         -\frac{10080-13952\nu+123\pi^2\nu+1440\nu^2}{128}\epsilon}\\
      \textcolor{blue}{-\frac{624\Delta\tilde{a}_-\tilde{a}_++24(1-8\nu)\tilde{a}_-^2-24(12\nu-61)\tilde{a}_+^2}{128}\epsilon}+\cdots\bigg]\frac{1}{2\ell^4}+\cdots\,.
   \end{multlined}
\end{aligned}
\eeq
where we introduced the reduced canonical orbital angular momentum as $\ell \equiv L/(GM\mu)$ (as opposed to $j$, to avoid confusion with the total angular momentum), and we dropped some terms in half-integer powers in $\epsilon$ which do not contribute to the map.\vskip 4pt

From the above expression we can then use,
\beq
\frac{\chi(J,\cE)+\chi(-J,\cE)}{2\pi} = \frac{\Delta\Phi(J,\cE)}{2\pi}\,,
\eeq
with $J$ the total angular momentum, to compute the periastron advance.
We colored in \textcolor{blue}{blue} the terms which are symmetric under $(L,a_\pm) \to (-L,-a_\pm)$, and therefore contribute (notice we factored out a $1/2$ in each one of them already), while the others cancel out. The result reads 
\beq
\begin{aligned}
  \frac{\Delta\Phi(\ell,a,\epsilon)}{2\pi} =
  &\left[\textcolor{blue}{3
  +\frac{3(2\nu-5)}{4}\epsilon
  +\frac{3(5-5\nu+4\nu^2)}{16}\epsilon^2}\right]\frac{1}{\ell^2}\\
  &\begin{multlined}
     +\bigg[\textcolor{blue}{-\frac{7\tilde{a}_++\Delta\tilde{a}_-}{2}
     -\frac{(\nu-6)\Delta\tilde{a}_-+(7\nu-18)\tilde{a}_+}{2}\epsilon}\\
     \textcolor{blue}{-\frac{3\left((15-14\nu+2\nu^2)\Delta\tilde{a}_-+(25-38\nu+14\nu^2)\tilde{a}_+\right)}{16}\epsilon^2}
       \bigg]\frac{1}{\ell^3}
   \end{multlined}\\
  &\begin{multlined}
    + \bigg[\textcolor{blue}{\frac{3(35+2\tilde{a}_+^2-10\nu)}{4}
         -\frac{10080-13952\nu+123\pi^2\nu+1440\nu^2}{128}\epsilon}  \\   
      \textcolor{blue}{-\frac{624\Delta\tilde{a}_-\tilde{a}_++24(1-8\nu)\tilde{a}_-^2-24(12\nu-61)\tilde{a}_+^2}{128}\epsilon}
      +\cdots
      \bigg]\frac{1}{\ell^4}+\cdots\,.
   \end{multlined}
\end{aligned}
\eeq
which neatly reproduces the value for the periastron advance derived in \cite{Sch12} to 3.5PN order.\footnote{
  We believe, however, there is a typo in the $C_Q$-independent contributions at ${\cal O}(|\cE|/L^4)$ in \cite{Sch12}.
  The term $\left(6-\frac{87}{4}\nu +3\nu^2\right)$ should be replaced by $\left(6-\frac{273}{8}\nu +3\nu^2\right)$, which readily gives a perfect match.}

\subsection{Periastron Advance to One-Loop}

Following \cite{Vines:2018gqi}, we can also use the map between test-body and two-body dynamics, to compute the periastron advance through \eqref{eq:chiphiS} to 2PM, and to all orders in velocity. Using the expressions in \cite{Vines:2018gqi}, we first reconstruct the two-body scattering angle via
\beq
\chi(m_1,a_1,m_2,a_2,v,b)=\frac{E}{M}\left(\frac{m_1}{M}\chi_t(M,a_1,a_2,v,b)+\frac{m_2}{M}\chi_t(M,a_2,a_1,v,b)\right)+\cdots\,.
\eeq
Hence, using our map,
\beq
\frac{\Delta\Phi(L,a_i)}{2\pi}=\frac{\chi(L,a_i)+\chi(-L,-a_i)}{2\pi}\,,
\eeq
together with the knowledge of the scattering angle in the test-body limit, we can compute the periastron advance to 2PM order. It is somewhat convenient to write the result either in terms of $b$ or $\hat L_{\rm cov} = L_{\rm cov}/\mu$, the covariant variables. These are related by $\hat L_{\rm cov} = \hat p_\infty b$, and to the canonical variables via \eqref{eqLb}.  The periastron advance thus reads
\beq
\begin{aligned}
  \frac{\Delta\Phi(L)}{2\pi G^2M^2} = &
  \frac{3(5\gamma^2-1)}{4{\hat L_{\rm cov}}^2\Gamma}
  -\frac{\gamma(5\gamma^2-3)(7a_++\Delta a_-)}{4{\hat L_{\rm cov}}^3\Gamma^2}\\
  &+\frac{3\left((1-6\gamma^2+5\gamma^4)(14\Delta a_-a_+-a_-^2)+(47-330\gamma^2+315\gamma^4)a_+^2\right)}{64{\hat L_{\rm cov}}^4\Gamma^3}\\
  = & \frac{3(4+v^2)}{4{\hat L_{\rm cov}}^2(1-v^2)\Gamma}
  -\frac{(2+3v^2)(7a_++\Delta a_-)}{4{\hat L_{\rm cov}}^3(1-v^2)^{\frac{3}{2}}\Gamma^2}\\
  &+\frac{3\left(v^2(4+v^2)(14\Delta a_-a_+-a_-^2)+(32+236v^2+47v^4)a_+^2\right)}{64{\hat L_{\rm cov}}^4(1-v^2)^2\Gamma^3}
\end{aligned}
\eeq
 in terms of $\gamma$ or the relative velocity, respectively. We found this representation much more simple, however, notice that there are spin-dependent terms {\it hidden} in the relationship \eqref{eqLb}, if written in terms of canonical variables. This expression reproduces our previous results in the overlapping regime of validity.

\section{Discussion}
 
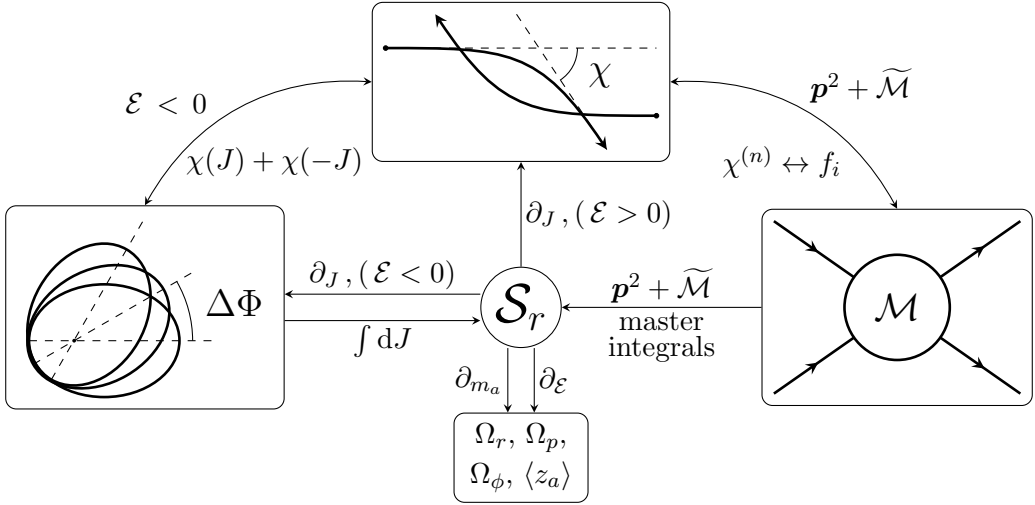
\begin{center}
  \begin{tikzpicture}
    \node[inner sep=4, rectangle, draw, rounded corners] (A) at (0:5) {
      \begin{tikzpicture}
        \draw[line width=1] (0,0) circle (0.7);
        \draw[line width=1,quark] (145:2) -- (145:0.7);
        \draw[line width=1,quark] (-145:2) -- (-145:0.7);
        \draw[line width=1,aquark] (35:2) -- (35:0.7);
        \draw[line width=1,aquark] (-35:2) -- (-35:0.7);
        \node at (0,0) {\Large$\cM$};
      \end{tikzpicture}
    };
    \node[inner sep=4, rectangle, draw, rounded corners] (B) at (90:3) {
      \begin{tikzpicture}
        [scale=0.3]
        \draw[dashed] (-6,1.5) -- (6,1.5);
	    \draw[dashed] (-0.2,3) -- (3.8,-3);
        \draw[->,line width=1] (-6,1.5) .. controls (-0.6,1.5) and (0.8,1.5) .. (3.8,-3);
        \draw[->,line width=1] (6,-1.5) .. controls (0.6,-1.5) and (-0.8,-1.5) .. (-3.8,3);
        \draw (2.5,1.5) arc (0:-57:1.7);
        \node at (3.5,0.3) {\Large$\chi$};
        \filldraw (-6,1.5) circle (0.1);
        \filldraw (6,-1.5) circle (0.1);
      \end{tikzpicture}
    };
    \node[inner sep=4, rectangle, draw, rounded corners] (C) at (180:5) {
      \begin{tikzpicture}
        [scale=0.25]
        \draw[line width=1] (0,0) ellipse (4 and 3);
        \draw[dashed] (-4,0) -- (6,0);
        \draw[line width=1,rotate=30] (0.1,0.8) ellipse (4 and 3);
        \draw[dashed,rotate=30] (-3.9,0.8) -- (6.1,0.8);
        \draw[line width=1,rotate=60] (0.8,1.4) ellipse (4 and 3);
        \draw[dashed,rotate=60] (-3.2,1.4) -- (6.8,1.4);
        \fill (-1.6,0) circle (0.1);
        \draw (4.7,0) arc (0:30:6.3);
        \node at (6.9,2) {\Large$\Delta\Phi$};
      \end{tikzpicture}
    };
    \node[draw,circle] (D) at (0,0) {\huge$\cS_r$};
    \node[inner sep=4, draw,rectangle, draw, rounded corners,align=center,text width=1.5cm] (E) at (-90:2) {$\Omega_r$, $\Omega_p$, $\Omega_\phi$, $\langle z_a \rangle$};
    \draw [<->] (A.north) to[bend right=32] node[below left,pos=0.3] {$\chi^{(n)}\leftrightarrow f_i$} node[above right=0.05] {$\bp^2+\widetilde\cM$} (B.east);
    \draw [->] (A.west) -- node[above] {$\bp^2+\widetilde\cM$}
  node[below,align=center,text width=2cm] {\baselineskip=10pt{master integrals\\}}(D);
    \draw [<->] (C.north) to[bend left=32]  node[above left=-0.2,align=center,text width=2.1cm] {$\cE<0$} node[below right=-0.1,pos=0.25] {$\chi(J)+\chi(-J)$} (B.west);
    \draw [->] (D.north) -- node[right] {$\partial_J\,,(\, \cE>0)$} (B.south);
    \draw [->] ([yshift=-5] C.east) -- node[below] {$\int\dd J$} ([yshift=-5] D.west);
    \draw [->] ([yshift=5] D.west) -- node[above] {$\partial_J\,,(\, \cE<0)$} ([yshift=5] C.east);
    \draw [->] ([xshift=5] D.south) -- node[right] {$\partial_\cE$} ([xshift=5] E.north);
    \draw [->] ([xshift=-5] D.south) -- node[left] {$\partial_{m_a}$} ([xshift=-5] E.north);
  \end{tikzpicture}
\end{center}
\subsection*{Conclusions}
In this paper we have developed further the boundary-to-bound dictionary introduced in paper\,I, relating scattering data to gravitational observables for bound states. Our main result in this paper is the existence of a remarkably simple relationship between the scattering angle and periastron advance in the conservative sector, 
\beq
\Delta\Phi(J,\cE)= \chi(J,\cE)+\chi(-J,\cE)\,, \qquad \cE<0\,, \label{eq:chiphid}
\eeq
obtained via analytic continuation in the angular momentum and binding energy. The above relationship allows us to reconstruct the (reduced) radial action directly from the scattering angle in the PM framework, yielding (for non-rotating bodies)
\beq
i_r(j,\cE) =  {\rm sg}(\hat p_\infty
)\chi^{(1)}_j(\cE) - j \left(1 + \frac{2}{\pi} \sum_{n=1}  \frac{\chi^{(2n)}_j({\cE})}{(1-2n)j^{2n}}\right)  \,,\label{eq:ird}
\eeq
with $j = GM \mu J$, the reduced orbital angular momentum, via analytic continuation to $\cE <0$. Using the expressions for the scattering angle as a function of the $f_n$'s to all orders derived in paper\,I, see \eqref{eq:phi2}, we have shown the equivalence of the above expression for the radial action with the one obtained in paper\,I in terms of master integrals, see \eqref{eq:irS}. This confirms the validity of the map between deflection angle and periastron advance to all PM orders. Moreover, the equivalence of representations allowed us to write a compact expression for the coefficients of the radial action, 
\beq
\beal
\label{eq:irgd}
  i_r (j,\cE) &= \frac{\hat p_\infty^2}{ \sqrt{-\hat p_\infty^2}} \frac{f_1}{2}
  + \frac{j}{2\sqrt{\pi}} \sum_{n=0}^\infty  \left(\frac{\hat p_\infty}{j}\right)^{2n}
  \Gamma\left(n-\frac{1}{2}\right)\sum_{\sigma\in\mathcal{P}(2n)}\frac{1}{\Gamma\left(1+ n-\Sigma^{\ell}\right)}\prod_{\ell} \frac{f_{\sigma_{\ell}}^{\sigma^{\ell}}}{\sigma^{\ell}!}\\
  &=  \frac{1}{2 \sqrt{-p_\infty^2}}\frac{{\widetilde \cM}_1}{M \mu} + \frac{j}{2\sqrt{\pi}} \sum_{n=0}^\infty \Bigg(  \frac{\Gamma\left(n-\frac{1}{2}\right)}{(\mu Mj)^{2n}}\,
  \sum_{\sigma\in\mathcal{P}(2n)}\frac{p_\infty^{2(n-\Sigma^{\ell})}}{\Gamma\left(1+n-\Sigma^{\ell}\right)}\prod_{\ell} \frac{{\widetilde \cM}_{\sigma_{\ell}}^{\sigma^{\ell}}} {\sigma^{\ell}!}\Bigg)\,,
\eeal
\eeq
to all PM orders, in terms of integer partitions of $2n=\sigma_\ell \sigma^\ell$, with $\Sigma^\ell = \sum_\ell \sigma^\ell$.
A partial resummation of one-loop terms can also be performed in closed-form, see \S\ref{sec:one}.
All of the gravitational observables for bound states follow from the above action via differentiation.
As an example we computed, in addition to the periastron advance, the azimuthal and radial frequency and redshift variable to two-loops, see \S\ref{sec:2loop}.
Agreement is found in the overlapping regime of validity of the PM and PN frameworks.
Yet, as discussed in paper\,I, the tree-level and one-loop  results also incorporate a series of exact-PN contributions, to all orders.
Moreover, as we argued here, the amplitude to three-loops will complete the knowledge of the $1/j^4$ corrections, e.g. to the periastron advance, to all orders in velocity.\vskip 4pt

Finally, via analytic continuation in the orbital angular momentum and spin, as well as the binding energy, we have shown that the relationship in \eqref{eq:chiphid} applies also once we include spin effects, provided we restricted the dynamics to aligned-spins in the direction of the angular momentum.
In that case, the periastron advance may be obtained from the deflection angle using \eqref{eq:chiphid}, with $J$ the canonical total angular momentum.
This implies that, in practice, we must flip the sign of the orbital angular momentum, $L \to -L$, as well as the spins, $a_i \to -a_i$.
Notice this implies the periastron advance is invariant under $J \to -J$, which is expected in the conservative sector.
As an example, using the results obtained in \cite{Vines:2018gqi} for the scattering angle to 3.5PN order, including spin effects, we have derived the periastron advance directly from \eqref{eq:chiphid}, and shown the agreement with the result obtained earlier in~\cite{Sch12}.
Finally, we have used the map between test- and two-body dynamics of \cite{Vines:2018gqi} to compute the periastron advance, including spin to one-loop order, to all orders in velocity.
We have also checked that the expression agrees with all the known limits.\vskip 4pt

There are, once again, many directions to continue exploring our dictionary further.
More pressing, perhaps, is the possibility to extend the impetus formula in \eqref{impetus} to spinning bodies.
There is also the intriguing connection between elementary particles and black holes, e.g. \cite{Vaidya:2014kza,Chung:2018kqs,Chung:2019duq,Arkani-Hamed:2019ymq,Guevara:2018wpp,Guevara:2019fsj,Guevara:2017csg,donalvines}.
We conclude our paper with a few words on these issues.

\subsection*{Impetus Formula \& Black Holes as Elementary Particles} 

The above manipulations strongly suggest that the impetus formula must remain valid, at least under some simplified conditions.
For starters, it is clear that for aligned-spins, the even-spin terms are spectators in the solution for $\bP^2$ from the Hamiltonian.
All we needed in paper\,I to demonstrate the impetus formula was the canonical representation of the linear momentum, together with the map to a non-relativistic quantum mechanical system~\cite{b2b1}.
Therefore, we expect the impetus formula to hold in such case.
Following the same steps, we conclude that the \emph{canonical} momentum is related to the scattering amplitude as in \eqref{impetus0},
\beq
\beal
\bP^2 (\bR,\cE,a_1,a_2) &= p_\infty^2 + \widetilde\cM^{S^{2k}}\big(\bR,p_{\infty},a_1,a_2\big)\,. \label{impetuss} \\
\eeal
\eeq 
where we denote the canonical position variable as $\bR$ here, to emphasize it's the coordinate associated with the canonical momentum. As before,
\beq
\widetilde\cM^{S^{2k}}(\bR,p_\infty,a_1,a_2) \equiv \frac{1}{2E} \int \frac{\dd^3\bq}{(2\pi)^3}\cM^{S^{2k}}(\bq,\bP,a_1,a_2)e^{i\bq\cdot \bR}\,,
\eeq
is the Fourier transform of the scattering amplitude normalized as in \eqref{impetus}, but involving --- in addition to spin-independent terms --- the even- and aligned-spins only.
Notice that, unlike before, the Fourier transform may produce a series of $1/R^\alpha$ terms at a given PM order, due to the coupling between spin and transfer momentum.\vskip 4pt

For example, let us consider the 1PM amplitude including only one of the particles carrying spin~$\ba$, which reads (recall $\mu M = m_1 m_2$) \cite{Guevara:2018wpp}
\beq
\cM^{S}_{\rm 1PM} (\bq,\bp,\ba) = 8\pi \frac{G\mu^2M^2}{\bq^2} \gamma^2 \sum_\pm (1\pm v)^2 e^{\pm i\bq\cdot \ba}\,.
\eeq
From this amplitude we only want to keep terms even in the spin, which means that the two factors of the velocity add up. Therefore, taking the Fourier transform and expanding, we find
\beq
\beal
\widetilde\cM^{S^{2k}}_{\rm 1PM} (\br,\cE,\ba) &=  \frac{2\mu^2(2\gamma^2-1)} {\Gamma}   \sum_{\ell}^{\rm even} \frac{1}{\ell!}\left ( ( i \ba)\cdot\boldsymbol\nabla\right)^\ell \frac{GM}{R}\\
&= \frac{2\mu^2(2\gamma^2-1)} {\Gamma}  \cos \left (\ba \cdot \boldsymbol\nabla\right) \frac{GM}{R} = \frac{2\mu^2(2\gamma^2-1)} {\Gamma}  \frac{GM r}{r^2+a^2\cos^2\theta}\,.
\eeal
\eeq
In the last step we followed the analysis in \cite{Vines:2016qwa} to re-write the answer in terms of new {\it oblate spheroidal} coordinates $(r,\theta)$ defined as
\beq
R\cos\Theta = r\cos\theta\,, \quad\quad R\sin\Theta = \sqrt{a^2+r^2}\sin\theta\,.
\eeq
 From here we obtain, 
\beq
\frac{\bP^2}{\mu^2}(\br,\cE,\ba) = \hat p_\infty^2(\cE) + \frac{2(2\gamma^2-1)} {\Gamma}  \frac{GM r}{r^2+a^2\cos^2\theta}\,.
\eeq
In the non-relativistic limit this becomes
\beq
\cE = \frac{\bP^2}{2\mu} + V +\cdots\,,
\eeq
with
\beq
V = - \frac{GM\mu\,r}{r^2+a^2\cos^2\theta}\,.
\eeq
The reader will immediately recognized the motion of a test-particle, $\mu$, in the potential produced by an object of mass $M$ with spin $\ba$, at linear order in $G$. This is --- yet another --- piece of evidence of the `elementary' nature of Kerr Black Holes.

\subsection*{Acknowledgements}

We thank the Munich Institute for Astro- and Particle Physics (MIAPP), supported by the DFG cluster of excellence ``Origin and Structure of the Universe'', and all the participants of the program ``Precision Gravity: From the LHC to LISA" for several fruitful discussions that led to paper\,I and the present paper. In~particular, Zvi Bern, Poul Damgaard, Guillaume Faye, Alfredo Guevara, Donal O'Connell, Radu Roiban, Chia-Hsien Shen, Mikhail Solon, Jan Steinhoff and Justin Vines. R.A.P would like to thank Justin Vines for many useful discussions on spin effects. R.A.P. acknowledges financial support from the ERC Consolidator Grant ``Precision Gravity: From the LHC to LISA"  provided by the European Research Council (ERC) under the European Union's H2020 research and innovation programme (grant agreement No. 817791), as well as from the Deutsche Forschungsgemeinschaft (DFG, German Research Foundation) under Germany's Excellence Strategy (EXC 2121) `Quantum Universe' (390833306).
  The research of G.K. is supported by the Swedish Research Council under grant 621-2014-5722, the Knut and Alice Wallenberg Foundation under grants KAW 2013.0235, 2018.0116, and 2018.0441, the Ragnar S\"{o}derberg Foundation (Swedish Foundations' Starting Grant), and in part by the US Department of Energy under contract DE--AC02--76SF00515.
\newpage
\appendix 
\section{Scattering Angle to
Periastron Advance to all PM orders} \label{appA}
In the following we show the (perturbative) equivalence of the radial action $\cS_r(J,\cE)$, as obtained in paper\,I in terms of master integrals, see \eqref{eq:irS},
\beq\label{eq:srOld}
\begin{aligned}
  i_r(J,\cE) &=- \sum_{n=0}^\infty \sum_{\sigma\in\cP(n)}
  \frac{(-1)^{\Sigma^{\ell}}\Gamma\left(\Sigma^{\ell} - \frac{1}{2}\right)}{2\sqrt{\pi}GM\mu}\cS_{\left\{n+2\Sigma^{\ell},\Sigma^{\ell}\right\}}(J,\cE) \prod_{\ell} \frac{D_{\sigma_{\ell}}^{\sigma^{\ell}}(\cE)}{\sigma^{\ell}!}\,, 
  \end{aligned}
\eeq
and the expression obtained in this paper in~\eqref{eq:irg}, reconstructed from the relationship between periastron advance and scattering angle,
\beq
\begin{aligned}
\label{eq:irgapp}
  i_r (j,\cE) &= \frac{\hat p_\infty^2}{ \sqrt{-\hat p_\infty^2}} \frac{f_1}{2}
  + \frac{j}{2\sqrt{\pi}} \sum_{n=0}^\infty  \left(\frac{\hat p_\infty}{j}\right)^{2n}
  \Gamma\left(n-\frac{1}{2}\right)\sum_{\sigma\in\mathcal{P}(2n)}\frac{1}{\Gamma\left(1+ n-\Sigma^{\ell}\right)}\prod_{\ell} \frac{f_{\sigma_{\ell}}^{\sigma^{\ell}}}{\sigma^{\ell}!}\,,
\end{aligned}
\eeq
both re-written here for the reader's convenience. The equivalence between the two guarantees that the PM expansion of the periastron advance, obtained from \eqref{partialJ}, is related via
\beq
\frac{\Delta\Phi^{(2n)}}{2\pi}=\frac{2}{\pi}\chi^{(2n)}_j\,,
\eeq
to the PM coefficients of the deflection angle, see \eqref{eq:pmjp} and \eqref{eq:pmangle}. At the same time, the representation in \eqref{eq:irgapp} provides a compact expression that can be used to derive all the gravitational observables for two-body bound states to any desired PM order.\vskip 4pt  
 
Let us start by staring at the master integrals, $\cS_{\{m,q\}}$ in \eqref{eq:srOld}, which can be written in terms of Hypergeometric functions:
\begin{align}
  \mathcal{S}_{\{2m,q\}} &= \begin{aligned}[t]
    &-i\,\delta_{m,0} (2q-1)B({\cal E}) A({\cal E})^{-q-\frac{1}{2}}\\
  	&+i \frac{(-1)^{m+q}A({\cal E})^{m-q}\Gamma\left(m-\frac{1}{2}\right)}
    {C(J,{\cal E})^{m-\frac{1}{2}}\Gamma(m-q+1)\Gamma\left(q-\frac{1}{2}\right)}
    {}_2F_1\left(m-\frac{1}{2},q-m;\frac{1}{2};\frac{B^2({\cal E})}{A({\cal E})C(J,{\cal E})}\right)\,,
  \end{aligned}\\
  \mathcal{S}_{\{2m+1,q\}} &= \begin{aligned}[t]
    &i\,\delta_{m,0}A({\cal E})^{\frac{1}{2}-q}\\
    &-2i\frac{(-1)^{m+q}A({\cal E})^{m-q}B({\cal E})\Gamma\left(m+\frac{1}{2}\right)}
    {C(J,{\cal E})^{m+\frac{1}{2}}\Gamma(m-q+1)\Gamma\left(q-\frac{1}{2}\right)}
    {}_2F_1\left(m+\frac{1}{2},q-m,\frac{3}{2};\frac{B^2({\cal E})}{A({\cal E})C(J,{\cal E})}\right)\,,
  \end{aligned}
\end{align}
where $A,B,C,D_n$ are defined in \eqref{abcd}. Notice that the angular momentum dependence enters only through $C(J,\cE)$. Using the power sum expansion of the Hypergeometric function, we can show that these can be re-expressed as
\begin{align}
  \mathcal{S}_{\{m,q\}} &= \begin{aligned}[t]
    &-i\,\delta_{m,0} (2q-1)B({\cal E}) A({\cal E})^{-q-\frac{1}{2}}\\
    &+\sum_{k~\mathrm{even}} \frac{(-1)^q i^{k+m+1}2^k\Gamma\left(\frac{1}{2}(m+k-1)\right)}{\Gamma(k+1)\Gamma\left(\frac{1}{2}(2+m-k-2q)\right)\Gamma\left(q-\frac{1}{2}\right)}\frac{A(\mathcal E)^{\frac{1}{2}(m-k-2q)}B(\mathcal E)^k}{C(J,\mathcal E)^{\frac{1}{2}(m+k-1)}}\,
  \end{aligned}\\
  \mathcal{S}_{\{m,q\}} &= \begin{aligned}[t]
    &i\,\delta_{m,1}A({\cal E})^{\frac{1}{2}-q}\\
    &+\sum_{k~\mathrm{odd}} \frac{(-1)^q i^{k+m+1}2^k\Gamma\left(\frac{1}{2}(m+k-1)\right)}{\Gamma(k+1)\Gamma\left(\frac{1}{2}(2+m-k-2q)\right)\Gamma\left(q-\frac{1}{2}\right)}\frac{A(\mathcal E)^{\frac{1}{2}(m-k-2q)}B(\mathcal E)^k}{C(J,\mathcal E)^{\frac{1}{2}(m+k-1)}}
  \end{aligned}
\end{align}
for $m$ even and odd respectively. We note that the summands are the same, but the corresponding sum goes over even and odd $k$ respectively.
In order to match the series expansion in $G/J$ of \eqref{eq:irg}, we expand the denominator involving $C(J,\cE)$ as follows
\beq
\begin{aligned}
  \frac{1}{C(\cE)^{\frac{1}{2}(m+k-1)}}
  &= \frac{1}{\left(J^2-G^2 M^2 p_\infty^2 f_2\right)^{\frac{m+k-1}{2}}}\\
  &= \sum_{s=0}^\infty \frac{\Gamma\left(\frac{1}{2}(2s+m+k-1)\right)}{\Gamma(s+1)\Gamma\left(\frac{1}{2}(m+k-1)\right)}\frac{\left((GM)^2 p_\infty^{2}f_2\right)^s}{J^{2s+m+k-1}}\,.
  \end{aligned}
\eeq
Plugging this expansion into eq.~\eqref{eq:srOld} we can (non-trivially) massage the radial action into the form
\beq\label{eq:irFromSr}
\begin{aligned}
  i_r = i_r^{(\infty)}-\sum_{n=0}^\infty\sum_{\sigma\in\cP(n)}\sum_k\sum_{s=0}^\infty
  &\frac{i^{2n-2k+2}}{2\sqrt{\pi}}\frac{\Gamma\left(\frac{1}{2}(n+k+2s+2\Sigma^\ell-1)\right)}{\Gamma\left(\frac{1}{2}(2+n-k)\right)}\\
  &\times \frac{ (\hat p_\infty^2)^{\Sigma^\ell+s+\frac{1}{2}(n+k)}}{j^{2s+k+n+2\Sigma^j-1}}\frac{f_1^k}{k!}\frac{f_2^s}{s!}\prod_{\ell} \frac{f_{2+\sigma_{\ell}}^{\sigma^{\ell}}}{\sigma^{\ell}!}\,,
\end{aligned}
\eeq
where the third sum is over even $k$ if $n$ is even and odd $k$ if $n$ is odd.
The leading term~$i_r^{(\infty)}$ represents the residue at $\infty$ in the contour integrals, see paper\,I. The proof continues by shifting $\sigma_{\ell}\rightarrow \sigma_{\ell}-2$ (and re-labeling $\ell\rightarrow \ell+2$) while identifying $k\rightarrow \sigma^1$ and $s\rightarrow \sigma^2$, such that we land the expression in~\eqref{eq:irg}, except for the first terms in both series.\footnote{Note that, in an intermediate step, the sums in eq.~\eqref{eq:irg} can be rewritten as
\beq
\sum_n \sum_{\sigma\in\cP(2n)} \rightarrow \sum_{\sigma^{\ell}}\,,
\eeq
with the condition that $\sigma_{\ell}\sigma^{\ell}$ is even.
This condition is most easily implemented by demanding that $k=\sigma^1$ is even or odd depending on the rest of the partition, leading to exactly the sum in \eqref{eq:irFromSr}.}
It is now a simple exercise to show the equality of $i_r^{(\infty)}$ with the leading term in \eqref{eq:irgapp}, almost by construction (see \S\ref{sec:rec}). This concludes the proof that the expression in \eqref{eq:irgapp} coincides with the PM expansion of  \eqref{eq:srOld}, to all orders.

\bibliographystyle{JHEP}
\bibliography{references2}

\end{document}